\documentclass[prd, aps, twocolumn, nofootinbib, showpacs, superscriptaddress]{revtex4-2}
\usepackage[utf8]{inputenc}
\usepackage{graphicx,mathtools,xcolor,latexsym,rotating,soul}
\usepackage[T1]{fontenc}
\usepackage{tikz}
\usepackage{comment}
\usepackage{amsmath, amssymb, amsthm, amsfonts,breqn}
\usepackage{multirow}
\usetikzlibrary{decorations.pathmorphing}
\usepackage[colorlinks=true,citecolor=green,urlcolor=blue,linkcolor=blue]{hyperref}
\usepackage{pifont}
\usepackage{gensymb}
\usepackage{enumitem}

\newtheoremstyle{named}{}{}{\itshape}{}{\bfseries}{.}{.5em}{\thmnote{#3 }#1}
\theoremstyle{named}

\newcommand*{\rom}[1]{\expandafter\@slowromancap\romannumeral #1@}

\newcommand\T{\rule[2.5 ex] {0pt}{0 ex}}       
\newcommand\B{\rule[-2.5 ex]{0pt}{0pt}}

\newcommand{\xmark}{\ding{55}}%

\makeatletter
\let\cat@comma@active\@empty
\makeatother
\begin{document}
\title{The non-relativistic limit of first-order relativistic viscous fluids}

\author{Abhishek Hegade K R}
\email{ah30@illinois.edu}
\affiliation{Illinois Center for Advanced Studies of the Universe, Department of Physics, University of Illinois at Urbana-Champaign, Urbana, IL 61801, USA}

\author{Justin L. Ripley}
\email{ripley@illinois.edu}
\affiliation{Illinois Center for Advanced Studies of the Universe, Department of Physics, University of Illinois at Urbana-Champaign, Urbana, IL 61801, USA}

\author{Nicol\'as Yunes}
\email{nyunes@illinois.edu}
\affiliation{Illinois Center for Advanced Studies of the Universe, Department of Physics, University of Illinois at Urbana-Champaign, Urbana, IL 61801, USA}
\begin{abstract}
Out-of-equilibrium effects may play an important role in the dynamics of neutron star mergers and in heavy-ion collisions. 
Bemfica, Disconzi, Noronha and Kovtun (BDNK) recently derived a causal, locally well-posed, and modally stable relativistic fluid model 
that incorporates the effects of viscosity and heat diffusion. 
We study the non-relativistic limit of this fluid model 
and 
show that causality for relativistic motion restricts the transport coefficients in the non-relativistic limit.
This restriction provides an upper bound on the ratio of the shear viscosity to the entropy density for 
fluids that can be described as relativistic within the BDNK model
and can be exactly modeled using the Navier-Stokes equation in the non-relativistic limit.
Furthermore, we show that the Fourier law of heat conduction must be modified 
by higher gradient corrections for such fluids.
We also show that the non-relativistic limit of the BDNK equations of motion are never hyperbolic, 
in contrast to the non-relativistic limit of extended variable models, whose non-relativistic equations of motion can be hyperbolic or not depending on the 
scaling of the transport coefficients present in the auxiliary equations for viscous degrees of freedom.
\end{abstract}
\maketitle
\section{Introduction}
Dissipative effects are expected to play an important role in relativistic fluid flow~\cite{Rezzolla-Book,Romatschke:2017ejr}. 
For non-relativistic fluid flow, the Navier-Stokes equations can successfully model dissipative effects, including viscosity and thermal conductivity~\cite{1961hhs..book.....C,1959flme.book.....L}.
In many situations, however, the flow speeds are highly relativistic, such as in heavy-ion collisions~\cite{2019NatPh..15.1113B,doi:10.1146/annurev-nucl-102212-170540}, or relativistic gravity cannot be ignored, such as during binary neutron star mergers~\cite{PhysRevLett.119.161101,2017ApJ...848L..13A}. 
The expected importance of dissipative effects in such situations is driving the development of relativistic fluid models that 
incorporate out-of-equilibrium effects, such as viscosity and thermal conduction.

Historically, several different approaches have been proposed to model 
relativistic viscous fluids~\cite{PhysRev.58.919,1959flme.book.....L,1976AnPhy.100..310I,1979RSPSA.365...43I,Muller-book}.
Following \cite{Rezzolla-Book,Bemfica:2020zjp}, 
we group these approaches into two classes: 
\textit{first-order fluid models} and 
\textit{extended-variable models}. 
In first-order models, 
viscous effects are modeled through the addition of first-order gradients
to the perfect fluid stress-energy tensor and the rest-mass current, 
whose covariant conservation leads to a relativistic version of the Navier-Stokes 
equations~\cite{PhysRev.58.919,1959flme.book.....L}.
Extended variable models (also called second-order, or Mueller-Israel-Stewart (MIS) models) build on the works of MIS~\cite{1976AnPhy.100..310I,1979RSPSA.365...43I,Muller-book},
by adding new degrees of freedom that are supposed to represent viscous 
corrections to the perfect fluid stress-energy tensor and rest-mass current. 
The conservation of this new stress-energy tensor and current, 
together with new dynamical equations for the new degrees of freedom (which can be inferred from 
kinetic theory in some cases), specify the evolution of the fluid flow.

Two of the earliest attempts to formulate relativistic viscous fluid models 
were proposed by Eckart~\cite{PhysRev.58.919} (``Eckart fluids''), 
and Landau and Lifshitz~\cite{1959flme.book.....L} (``Landau fluids''). 
Both Eckart and Landau fluids are examples of first-order fluid models. 
The main difference between the two models is that for Landau fluids,  
the rest-mass current receives viscous corrections, in addition to the stress-energy tensor, while in Eckart models, only
the stress-energy tensor recieves viscous corrections.
Although the conservation of these models' stress-energy tensor and rest-mass current
lead to intuitive, relativistic generalizations of the Navier-Stokes equations, 
they suffer from several severe problems. 
The most important of these are that the equations of motion are modally unstable about 
thermal equilibrium in flat (Minkowski) spacetime 
and that they are acausal in the relativistic regime 
~\cite{1983AnPhy.151..466H,1985PhRvD..31..725H}.

These problems led many researchers to discard first-order relativistic fluid models and to 
instead concentrate on extended variable models. 
Some of the latter showed promise because they were shown to lead to hyperbolic equations and 
to have modally stable equilibrium solutions in the \textit{linear} regime \cite{1983AnPhy.151..466H,OLSON199018}.
Several modern versions of the extended variable models have been used to describe 
relativistic fluid flows observed in heavy-ion collisions~\cite{Baier:2007ix,Denicol_2012}.
Despite their widespread use in numerical simulations, however,
\emph{no} extended variable model has been shown to be causal, stable, and to have a locally
well-posed initial value problem without making a set of 
restrictive assumptions on the structure of the equations (for more detail see,~\cite{Bemfica:2019cop,Bemfica:2020xym}).
Moreover, the introduction of extra parameters in these 
models makes them less appealing to some from a physical point of view.

Recently, attention has shifted back to first-order models because a 
generalization of the first-order Eckart fluid model, derived by Bemfica, Disconzi, and Noronha~\cite{Bemfica_2018,Bemfica:2020zjp} and Kovtun~\cite{Kovtun_2019,Hoult:2020eho}, has been shown to 
circumvent the above mentioned shortcomings of extended variable models.
Following earlier work (e.g.~\cite{Pandya:2021ief}), 
we shall refer to this fluid framework as the \emph{BDNK fluid model}, or as
a \emph{BDNK fluid} for short.
BDNK fluids are modally stable around thermal equilibrium in Minkowski spacetime
(\cite{Bemfica:2020zjp,PhysRevD.98.104064,PhysRevD.100.104020,Hoult_2020}).
Equally important, the model is causal and has a locally well-posed initial value problem 
(the theory is strongly hyperbolic), even for non-equilibrium solutions \cite{Bemfica:2020zjp}. 
These properties hold provided certain causality and stability conditions are satisfied 
(see App.~A and B of~\cite{Bemfica:2020zjp}).
The key insight of BDNK was that the problems of causality, stability, and hyperbolicity 
can be resolved by using a different \emph{hydrodynamic frame} 
than those used by Eckart and by Landau and Lifshitz.
A hydrodynamic frame 
(not to be confused with geometric notions of frame, such as a coordinate frame or a tetrad) 
is defined by how one chooses to define the equilibrium fluid variables 
when they are no longer in thermal equilibrium \cite{Rezzolla-Book}.
More concretely, a hydrodynamic frame can be thought of as a choice of parameterization
of the stress-energy tensor in terms of physical quantities, 
such as the energy density, the flow velocity and the particle number density.
BDNK~\cite{Bemfica:2020zjp,Kovtun_2019} 
showed that certain choices of hydrodynamic frame allow for fluid equations of motion 
(that arise from the conservation of the stress-energy tensor and the rest-mass current of first-order models) 
to be causal, modally stable (about thermal equilibrium, in flat spacetime), 
and to have a locally well-posed initial value problem.

Specifically, BDNK fluids are characterized by the addition of transport coefficients that 
enter the viscous generalization of the perfect fluid stress-energy tensor. 
Three of these are the thermal conductivity, 
the shear viscosity and the bulk viscosity of the fluid, 
which also appear in the Navier-Stokes equations and have a clear physical interpretation.
In addition to these, BDNK fluids have several additional transport coefficients 
whose physical role and relevance is less clear.
In the language of effective field theories, 
these additional transport 
coefficients---which are partly introduced to ensure causality---parameterize  
``high-energy'' or ``ultraviolet'' physics (for a review, see \cite{Burgess:2007pt}).
The transport coefficients in the BDNK model must satisfy a set of inequalities
in order for the theory to be causal and modally stable~\cite{Bemfica:2020zjp}\footnote{We note
that in \cite{Bemfica:2020zjp}, the authors considered only a subclass of all possible
first-order, hyperbolic theories.  
We emphasize that by ``BDNK fluid'', we mean the class of fluid models where the 
fluid stress-energy tensor receives the most general set of first-order gradient corrections, but the fluid current receives no gradient
corrections; see Eq.~\eqref{eq:parameterization-general}. For more discussion
see also \cite{Kovtun_2019}.}.

A major challenge to both understanding the physical predictions of BDNK fluids, in addition
to numerically evolving their equations of motion, lies in finding a fluid frame that 
satisfies the causality and modal stability constraints. 
Very few frames have been found, and presently only one frame has been
numerically evolved, and been compared to predictions of extended variable models 
~\cite{Pandya:2021ief,Pandya:2022pif,Pandya:2022sff}. With an eye towards finding physically acceptable BDNK fluid frames, 
our goal in this paper is to understand 
the implications of the BDNK causality constraints for non-relativistic BDNK fluid flow,
where the physical interpretation of the fluid motion may be more readily apparent. 
More specifically, in this paper we answer the following questions:
\begin{enumerate}
    \item[(1)] What is the non-relativistic limit
    (i.e. the limit where the speed of light $c\to\infty$) 
    of the fluid equations that arise from the BDNK fluid model? 
    Under what conditions do these non-relativistic fluid equations reduce to the Navier-Stokes system?
    \item[(2)] How do the causality constraints impact the behavior of the BDNK model 
    for non-relativistic viscous flows?
    \item[(3)] How is the non-relativistic limit of the BDNK model different from the non-relativistic limit of extended variable models? 
\end{enumerate}
As we shall see, the answers to these questions will lead to constraints on the physical transport coefficients of the BDNK model and illustrate the difference between the BDNK model and extended variable models. The answers to these questions are also important to understand the applicability of the BDNK model to astrophysically relevant scenarios where out-of-equilibrium effects might be important, such as in the dynamics of neutron stars, accretion disks, and white dwarfs.

To analyze the answers to these questions, we make two physically reasonable assumptions:
(i) the non-relativistic limit of the BDNK fluid must reduce to a system of 
hydro-dynamical equations with non-zero viscosity and heat conduction, and 
(ii) the BDNK fluid retains the same scaling in powers of $c^{-1}$ up to $\mathcal{O}(c^{-5})$ as that of a perfect fluid, so that the structure of the post-Newtonian metric does not change when one includes out-of-equilibrium effects.

Concerning question (1),
we first derive the non-relativistic equations of the BDNK model and investigate the hyperbolicity of the non-relativistic equations.  
We find that these equations are \emph{not} hyperbolic.
This result by itself is not too surprising, as for example general relativity
(a hyperbolic, relativistic theory) reduces to Newtonian gravity (an elliptic theory)
in the non-relativistic limit.
In the context of relativistic viscous fluid theories though, this result may be less obvious.
For example, the non-relativistic limit of the equations of motion for
extended-variable models can reduce to either hyperbolic or non-hyperbolic theories,
depending on the scaling of the parameters present in the model (see Sec.~\ref{sec:MIS-model}.)
Finally, we also show that if some of the BDNK transport coefficients scale as $c^{-2}$, 
the non-relativistic equations of motion can reduce to the Navier-Stokes equations. Moreover, we also show that, in the BDNK fluid model, the non-relativistic Fourier-law for heat conduction is always 
modified by higher-order gradient corrections, 
due to the presence of the \emph{relativistic} causality constraints.

With regard to question (2), 
the structure of relativistic causality constraints 
involve both the physical transport coefficients 
(such as the bulk viscosity, shear viscosity and thermal conductivity), 
and additional transport coefficients, which should
not be large in the non-relativistic limit.
Using this, we show that the non-relativistic limit of the causality constrains
can bound the values 
of the non-relativistic limit of the physical transport coefficients.
In other words, the relativistic causality constraints
affect the properties of the \emph{non-relativistic} limit of the equations of motion.

More specifically, 
we show that
the causality constraints restrict the ratio of the thermal conductivity to 
the kinematic viscosity for non-relativistic fluid flows, provided either that
(a) the non-relativistic equations of motion reduce to the Navier-Stokes equations,
or 
(b) the dissipation timescales set by the new BDNK transport coefficients are much smaller
than those set by the viscosity and heat transport coefficients.
This inequality can be used to provide an \textit{upper} 
bound on the ratio of the shear viscosity to the entropy density
in terms of the thermal conductivity of the fluid (see Eq.~\eqref{eq:upper-bound-eta-by-s}).
This upper bound is satisfied by theoretical results available from Chapman-Enskog theory.
Our results also complements 
the Kovtun-Son-Starinets (KSS)~\cite{Kovtun:2004de} 
conjecture derived from holography, which provides a lower bound of the ratio of shear viscosity to entropy density.
Finally, we show that one can combine our inequality with the non-relativistic limit of the 
KSS inequality to obtain a lower bound on the product of the ratio of the thermal conductivity to the square of the entropy density and the density of the fluid.

With regard to question (3), we analyze the non-relativistic limit of an extended-variable model \cite{Bemfica:2019cop}.
We show that the extended variable model can admit both hyperbolic and non-hyperbolic
non-relativistic equations depending on the scaling of the transport coefficients with powers of $c^{-1}$.
This happens because the extended variable models have additional equations for the viscous degrees of freedom (apart from the stress energy conservation equations).
This is in contrast to the BDNK fluid model, whose non-relativistic limit is always not hyperbolic.
We also derive limits on the value of the bulk viscosity of this extended variable model by studying the non-relativistic limit.

A detailed investigation of the answers to any of these questions had not yet been performed in previous work on the BDNK model.
Some partial results regarding question (2) outside of conformal fluids do exist in the context of numerical simulations~\cite{Pandya:2022pif}.
Although the frame presented in~\cite{Pandya:2022pif} satisfies the mathematical constraints derived by BDNK~\cite{Bemfica:2020zjp},
we will show here that that frame violates
conditions (i) (i.e.~that the non-relativistic equations of motion have nonzero viscosity)
and (ii) (i.e.~the preservation of the structure of the post-Newtonian metric). 
So, new frames need to be investigated before one can use
the BDNK model to study astro-physically relevant systems. 
Our detailed investigation provides constraints on the physical transport coefficients that appear in the BDNK fluid model,
which may help in finding new fluid frames.

The remainder of this paper presents the details that led to the conclusions summarized above and is organized as follows.
We consider several toy models to illustrate the possible types of behavior possible 
in the non-relativistic limit in Sec.~\ref{sec:toy-models}.
We introduce the BDNK fluid model and show that the non-relativistic BDNK equations of motion
are not hyperbolic
and derive several constraints on the transport coefficients in Sec.~\ref{sec:First-Order-Models}.
Next, we analyze a simple extended variable model in Sec.~\ref{sec:MIS-model},
and show that the equations of motion can be either hyperbolic or not, depending
on how the transport coefficients scale with the speed of light $c$.
Finally, we present our conclusions in Sec.~\ref{sec:conclusions}.
We work in SI units and keep factors of $G$ and $c$ in all our expressions to make
our non-relativistic expansion clear. We also work with the metric signature $(-,+,+,+)$.

\section{Possible types of behavior in the non-relativistic limit}\label{sec:toy-models}

For the sake of completeness, 
before performing the non-relativistic expansion of the BDNK equations of motion, 
we first consider a simplified toy-model, 
which captures many of the main points and subtleties of the non-relativistic expansion of the BDNK model.
We perform the non-relativistic expansion of a simple model in Sec.~\ref{subsec:NR-limit-meaning},
and present the possible types of non-relativistic behavior in Sec.~\ref{subsec:classification-NR}.
Readers already acquainted with post-Newtonian theory can directly go to the next section.
\subsection{Meaning and validity of the non-relativistic limit}\label{subsec:NR-limit-meaning}
To illustrate the set of assumptions that go into the derivation of a non-relativistic limit of a
set of equations, we first consider the non-relativistic limit 
for a simplified model for relativistic diffusion.
We introduce a scalar field $\phi$ 
that satisfies the following conservation equation for a current $B^{\alpha}$
\begin{subequations}
\label{eq:full-B-alpha-1-eqns}
\begin{align}\label{eq:B-alpha-1}
    B^{\alpha} 
    &\equiv 
    \phi u^{\alpha}  
    - 
    \tau_2 \nabla^{\alpha }\phi 
    \,,\\
    \label{eq:eom-toy-model-1}
    \nabla_{\alpha} B^{\alpha}
    &=
    0
    \,,
\end{align}
\end{subequations}
where $u^{\alpha}$ is a constant 4-velocity vector that advects the scalar field and 
$\tau_2$ is a coefficient of kinematic viscosity, which recall has units of length squared divided by time (or units of length in geometric units).  

We now set up some notation used in the non-relativistic expansion.
Given a local Lorentz frame $x^{\alpha} = (ct, x, y,z)$, the line element is given by
\begin{equation}
    -c^2 d\tau^2 = \eta_{\alpha \beta} dx^{\alpha} dx^{\beta}\,, 
\end{equation}
where $\eta_{\alpha \beta}= \text{diag} (-1,1,1,1)$ is the Minkowski metric in these coordinates.
The 4-velocity is normalized to $u^{\alpha} u_{\alpha} = -c^2$,
which implies that $u^{\alpha} = \left(cW, W v^i \right)$, where
\begin{equation}
     W 
    \equiv 
    \frac{1}{\sqrt{1 - c^{-2}v_iv^i}}
    \,.
\end{equation}
For simplicity, we compute the non-relativistic limit for 
planar fluid flow in the $x$ direction with 4-velocity
\begin{align}
    u^{\alpha} 
    = 
    W\, \left(c, v,0,0 \right)\,.
\end{align}

Assuming that $\tau_2$ and $v$ are constant,
the equations of motion [Eq.~\eqref{eq:eom-toy-model-1}] can then be  
decomposed in a local Lorentz frame as 
\begin{align}
    \label{eq:local_lorenz_frame_advection_diffusion}
    \frac{\tau_2}{c^2} \partial_t^2 \phi 
    -
    \tau_2 \partial_x^2 \phi 
    + 
    W\partial_t \phi 
    + 
    W v
    \, 
    \partial_x \phi 
    &= 
    0\,.
\end{align}
The non-relativistic expansion of the solutions to the above equation of motion is one in which the characteristic velocities are assumed to be much smaller than the speed of light. More precisely, one assumes that the system possesses a characteristic timescale ($t_c$) and a characteristic length scale ($x_c$), such that the characteristic velocity $v_c = x_c/t_c$ is much smaller than the speed of light $v_c/c \ll 1$. 
In particular, this implies that the $N$th time derivative acting on the field $\phi$ is much smaller than the $N$th spatial derivative times the speed of light to the Nth power, e.g., $\partial_t \phi \ll c \partial_x \phi$. 
Moreover, when there is a characteristic internal timescale $t_{\text{int}}$ in the problem, one must also assume that $t_c \gg t_{\text{int}}$.
For the example presented above, $t_{\text{int}} =\tau/{c^2}$ so we assume that the characteristic timescale of the problem is much longer than the characteristic relaxation time of the problem $t_c \gg \tau/{c^2} $.
The orders of a non-relativistic expansion are colloquially labeled as powers of $1/c$, where one implicitly means powers of $v_c/c$ and $t_{\text{int}}/t_c$. 
The spacetime regions where the non-relativistic expansion holds is called the near-zone in post-Newtonian theory.

When these conditions hold, we can expand the equations of motion in the non-relativistic limit, i.e.~to leading order in the non-relativistic expansion.
For example, to leading order in this expansion,
\begin{enumerate}
    \item the Lorenz factor $W$ can be set to unity.
    \item the wave retardation effect due to $({\tau_2}/{c^2}) \partial_t^2 \phi$ can be set to zero.
\end{enumerate}
With this, the relativistic advection-diffusion equation 
\eqref{eq:local_lorenz_frame_advection_diffusion} to leading order in the non-relativistic expansion is
\begin{align}
    \label{eq:toy_nr_limit}
    \partial_t \phi 
    + 
    v\, \partial_x \phi
    -
    \tau_2 \partial_x^2 \phi 
    = 
    \mathcal{O}\left(c^{-2}\right)
    \,.
\end{align} 
We see that Eq.~\eqref{eq:toy_nr_limit} takes the form of the advection-diffusion equation.

\subsection{Classification of the non-relativistic behavior}
\label{subsec:classification-NR}

The previous example illustrates that a causal and hyperbolic theory might reduce to an acausal theory of advection-diffusion type in the non-relativistic regime.
In general, there are three different types of behavior possible for a relativistic, hyperbolic
theory, in the non-relativistic limit:
\begin{enumerate}
    \item The non-relativistic limit of the equations of motion are never hyperbolic.
    \item The non-relativistic limit of the equations of motion are always hyperbolic.
    \item The non-relativistic limit of the equations of motion could be hyperbolic or not, depending on the scaling of $1/c$ of the parameters present in the theory.
\end{enumerate}

Several simple examples exist of theories that exhibit type (1) behavior, and thus, their equations of motion are never hyperbolic in the non-relativistic limit. One simple example is the relativistic wave equation for a scalar field $A$,
\begin{align}
    \Box A  
    = 
    -
    \frac{1}{c^2} \partial_t^2 A 
    + 
    \partial_j \partial^j A 
    = 0 
    \,.
\end{align}
In the non-relativistic limit, the equations of motion reduce to an elliptic equation
\begin{align}
    \partial_j \partial^j A = 0\,,
\end{align}
which physically implies that information propagates almost instantaneously in the non-relativistic regime. Another example is general relativity, whose field equations reduce to Newton's in the non-relativistic limit~\cite{Poisson-Will}. The equation of motion in Newton's theory of gravity is the Poisson equation, which is also elliptic.
In Sec.~\ref{sec:First-Order-Models}, we will show that the non-relativistic BDNK fluids equations of motion are not hyperbolic.

Examples also exist of theories that exhibit type 2 behavior, yielding equations of motion that remain hyperbolic in the non-relativistic limit. One example is scalar field in which the scalar 
$\phi$ is advected along a constant timelike four velocity vector $u^{\mu}$. The equation of motion of this theory is 
\begin{equation}
\begin{aligned}
    \nabla_{\mu} \left(\phi u^{\mu} \right)
    &=
    0\,,\\ 
    u^{\alpha}\partial_{\alpha} 
    & = 
    \frac{1}{\sqrt{1 - v_i v^i c^{-2}}}\partial_t
    +
    \frac{v^j}{\sqrt{1 - v_i v^i c^{-2}}}
    \partial_j
    \,,
\end{aligned}
\end{equation}
and its non-relativistic limit 
is the non-relativistic advection equation
\begin{align}
    \partial_t \phi + \partial_j (v^j \phi) 
    &=0
    \,.
\end{align}
Since this equation does not have any heat diffusion (compare e.g.~with Eq.~\eqref{eq:toy_nr_limit}), we see that it is always hyperbolic. 
Another example of theory of type 2 is the theory of relativistic perfect fluids.
The equations of motion
for perfect fluids are hyperbolic (in both relativistic and non-relativistic regime) given a suitable choice of an equation of state. 

Finally, let us provide an example of a theory whose non-relativistic equations
of motion could be hyperbolic or non-hyperbolic.
Consider a slightly more general current than the one of Eq.~\eqref{eq:B-alpha-1} with a standard conservation equation, namely
\begin{subequations}
\begin{align}
    \hat{B}^{\alpha}  
    &\equiv 
    \phi u^{\alpha} 
    +
    \tau_1 u^{\alpha} \nabla_{\mu} \left(\phi u^{\mu} \right)
    - 
    \tau_2 \nabla^{\alpha} \phi 
    \,,\\ 
    \nabla^{\alpha} \hat{B}_{\alpha} 
    &=
    0\,.
\end{align}
\end{subequations}
The principal part of the conservation equation is
\begin{align}
    \label{eq:principal-symbol-bhat}
    \nabla_{\alpha}\hat{B}^{\alpha}
    &= 
    \left(
        \tau_1 u^{\alpha}u^{\beta} 
        - 
        \tau_2 \eta^{\alpha \beta}
    \right) \nabla_{\alpha}\nabla_{\beta} \phi 
    +
    \mathrm{l.o.t.}
    ,
\end{align}
where ``l.o.t.'' denotes lower-order (in derivatives) terms.
Examining the principal part of these relativistic equations, we conclude that the theory is hyperbolic and causal when
\begin{align}\label{eq:causality-toy-model}
    0
    <
    \frac{\tau_2}{\tau_1 c^2 + \tau_2}
    \leq 
    1
    \,,
\end{align}
which can be derived from Eq.~\eqref{eq:principal-symbol-bhat} in the rest-frame of the fluid ($v=0$)~\cite{Rezzolla-Book}.
To find the non-relativistic limit of $\nabla_{\alpha}\hat{B}^{\alpha}=0$, 
we expand the transport coefficients as
\begin{subequations}\label{eq:tau-1-2-scaling}
\begin{align}
    \tau_1 
    &= 
    \tau_1^{(0)} + \frac{\tau_1^{(2)}}{c^2} + \mathcal{O}{(c^{-4})} 
    \,, \\
    \tau_2 
    &= 
    \tau_2^{(0)} + \frac{\tau_2^{(2)}}{c^2} + \mathcal{O}{(c^{-4})}
    \,.
\end{align}
\end{subequations}
Assuming $(\tau_1,\tau_2)$ are constant, the non-relativistic limit of the equations of motion is then
\begin{align}
    \tau_1^{(0)} \left(\partial_t^2 + 2 v^j \partial_j \partial_t + v^i v^j \partial^2_{ij} \right) \phi 
    + 
    \partial_t \phi + \partial_j (v^j \phi) 
    \nonumber \\
    = 
    \tau_2^{(0)} \partial_j \partial^j \phi\,.
\end{align}
When $\tau_1^{(0)} \tau_2^{(0)} >0$, these equations can be shown to be hyperbolic.
When $\tau_1^{(0)} = 0$, one finds the advection-diffusion equation, which is parabolic. 
Depending on the leading-order behavior of $\tau_{1,2}$ in the non-relativistic limit (i.e.,~depending on the values of $(\tau_1^{(0)}, \tau_2^{(0)})$), 
the non-relativistic limit of the conservation equation 
$\nabla_{\alpha} \hat{B}^{\alpha}$ can be either  hyperbolic or parabolic.

Does the above fact (i.e.~that the non-relativistic limit of this theory leads to parabolic equations when $\tau_1^{(0)} = 0$) lead to problems related to causality for the relativistic theory? The answer is clearly no.
To see this, we expand Eq.~\eqref{eq:causality-toy-model} using Eq.~\eqref{eq:tau-1-2-scaling}
to see that the relativistic theory remains causal when
\begin{align}\label{eq:causality-NR-toy-model}
    \frac{\tau_2}{c^2 \tau_1 + \tau_2} 
    = 
    \frac{\tau_2^{(0)}}{\tau_2^{(0)} + \tau_1^{(2)}} 
    + 
    \mathcal{O}(c^{-2}) \leq 1
    \,.
\end{align}
or simply when
\begin{equation}
\frac{\tau_1^{(2)}}{\tau_2^{(0)}} >0\,.
\label{eq:example-constraint}
\end{equation}
%
What we have then found is that this relativistic theory is causal even when $\tau_1^{(0)} = 0$ provided that Eq.~\eqref{eq:example-constraint} holds, but its non-relativistic limit becomes acausal in the regimes of spacetime where this limit is valid (i.e.,~in the near zone). 
This apparent contradiction is an artifact of the truncation of the non-relativistic expansion when staying at leading order; it would be resolved if, instead of taking the non-relativistic \textit{limit}, one were to 
study the relativistic equations.
Equation~\eqref{eq:example-constraint} is then a constraint on the relativistic theory, and constraints of this kind will be used in Sec.~\ref{sec:First-Order-Models}, and
in Sec.~\ref{sec:MIS-model}.

\section{The BDNK fluid model}\label{sec:First-Order-Models}
In this section, we present the stress energy tensor and the field equations for the BDNK fluid model in Sec.~\ref{subsec:Field-equation-BDNK} and then we present the causality constraints for this model in Sec.~\ref{subsec:Causality-BDNK-general}.
\begin{table*}[t]
    \caption{Comparison of notation between BDN~\cite{Bemfica:2020zjp}, 
    and the textbook references \cite{Rezzolla-Book}, \cite{Poisson-Will}. 
    As we work with chemical potentials when
    determining the equation of state, 
    we adopt the notation used in \cite{Rezzolla-Book} 
    for the fluid variables, and reserve the
    use of $\mu$ for the chemical potential.
    A complete summary of our notation is provided in Appendix~\ref{appendix:units}.}
    \centering
    \begin{tabular}{|c|c|c|c|}
        \hline
       Physical Quantity & 
       Rezzolla and Zanotti \cite{Rezzolla-Book} & 
       Poisson and Will \cite{Poisson-Will} & 
       BDNK \cite{Bemfica:2020zjp}  \T\B 
       \\
        \hline
        Rest-mass Density & 
        $\rho$ & 
        $\rho$ & 
        $m_b n$  \T\B 
        \\
        \hline
        Total Energy Density & 
        $e = \rho c^2 (1 + \epsilon\, c^{-2})$ & 
        $\mu = \rho c^2 + \epsilon$ & 
        $\varepsilon$ \T \B 
        \\ 
        \hline
        Pressure & 
        $p$ & 
        $p$ & 
        $P$  \T \B 
        \\
        \hline
    \end{tabular}
    \label{tab:RZ-BDNK}
\end{table*}
\subsection{Field Equations}\label{subsec:Field-equation-BDNK}
Let us begin by reviewing the first-order relativistic viscous fluid models introduced by 
BDNK~\cite{Kovtun_2019,Bemfica:2020zjp}.
Our notation for the fluid variables follows~\cite{Rezzolla-Book},
which we emphasize is \textit{not} the notation followed by BDNK.
Table~\ref{tab:RZ-BDNK} summarizes the differences between our notation and other common
definitions for the fluid variables.
Following~\cite{Rezzolla-Book}, we denote the baryon number density by $n$, 
the rest-mass density by $\rho$, the total energy density by $e$, the pressure by $p$ 
and the specific internal energy by $\epsilon$.
A complete summary of our notation is provided in Appendix~\ref{appendix:units} and in Table~\ref{tab:units}.

Consider a fluid moving with a four-velocity $u^{\mu}$.
We decompose the stress-energy tensor $T_{\mu\nu}$ and the rest-mass current 
$J^{\mu}$ into components parallel and perpendicular to $u^{\mu}$:
\begin{align}
    \label{eq:decomp_tmunu}
    T_{\mu\nu} 
    &= 
    \mathcal{E} u_{\mu} u_{\nu}
    +
    \mathcal{P} \Delta_{\mu \nu}
    +
    2\mathcal{Q}_{(\mu}u_{\nu)}
    +
    \mathcal{T}_{\mu\nu}
    \,, \\
    \label{eq:J-decomp}
    J^{\mu} 
    &= 
    \mathcal{N} u^{\mu} 
    + 
    \mathcal{J}^{\mu}
    \,,
\end{align}
where, $\Delta_{\mu \nu} \equiv g_{\mu\nu} + c^{-2} u_{\mu} u_{\nu}$ 
is the projection tensor.
The quantities $\mathcal{E}$, $\mathcal{P}$, $\mathcal{Q}^{\mu}$, and $\mathcal{T}^{\mu\nu}$
can be respectively interpreted as the generalized energy density, pressure, 
heat flux, and shear density.
For a relativistic perfect fluid, the only nonzero terms are
\begin{align}
    \mathcal{E}_{\rm PF} &= \frac{e}{c^2}\,, \\
    \mathcal{P}_{\rm PF} &= p \,.
\end{align}
For non-perfect fluids, the definitions of $\mathcal{E}$, $\mathcal{P}$, $\mathcal{Q}^{\mu}$ 
and $\mathcal{T}^{\mu\nu}$ are less obvious. 
Some common choices based on physical intuition can fail; for example, the definitions 
of Landau and Lifshitz~\cite{1959flme.book.....L}, and of Eckart~\cite{PhysRev.58.919} lead to equations of motion that fail to be causal and whose solutions can develop exponentially-growing modes in flat spacetime
\cite{1983AnPhy.151..466H,1985PhRvD..31..725H}.

In the BDNK approach, non-equilibrium effects are modeled through the addition of
gradients of the fluid fields to the stress-energy tensor and fluid current. 
The coefficients of this first-order gradient expansion are then chosen so that
the conservation laws lead to hyperbolic differential equations, equilibrium
solutions are linearly stable, and that in thermal equilibrium the stress-energy
tensor reduces to that of a perfect fluid \cite{Kovtun_2019,Bemfica:2020zjp}.
Presently, the arguments provided in \cite{Bemfica:2020zjp} to rigorously prove
the strong-hyperbolicity of the equations of motion can only be applied when the
rest-mass current receives no gradient corrections, so we do not consider gradient
corrections to that term. 
Provided there are no non-equilibrium
fluctuations to the rest-mass energy (that is, provided non-equilibrium effects do
not destroy and/or create particles), little physical explanatory power should
be lost by making this restriction\footnote{While this assumption should be true for the baryonic matter of neutron stars, it does not hold for the
quark-gluon plasma made during the collision of heavy nuclei.}.

The most general stress energy tensor with zero mass diffusion current ($\mathcal{J}^{\mu} = 0$) and no corrections to the rest-mass energy density $\rho$ is 
is given by
\begin{subequations}\label{eq:parameterization-general}
\begin{align} 
\label{eq:E-def-general-param}
    \mathcal{E} 
    \equiv& 
    \frac{e}{c^2} + \frac{\tau_{\epsilon,1}}{c^2} u^{\alpha} \nabla_{\alpha} \epsilon 
    +
    \frac{\tau_{\epsilon,2}}{c^2} \theta
    \nonumber \\
    &+
    \frac{\tau_{\epsilon,3}}{c^4} u^{\alpha}\nabla_{\alpha} \left(\frac{\mu}{T}\right)
    \,, \\
\label{eq:P-def-general-param}
    \mathcal{P}
    \equiv& 
    p 
    +
    \tau_{p,1} u^{\alpha} \nabla_{\alpha} \epsilon
    -
    \zeta \theta
    +
    \tau_{p,2} \theta
    \nonumber\\
    &
    +
    \frac{\tau_{p,3}}{c^2} u^{\alpha}\nabla_{\alpha} \left(\frac{\mu}{T}\right) 
    \,,\\
\label{eq:Q-def-general-param}
    \mathcal{Q}^{\mu} 
    \equiv& 
    \frac{\tau_{Q,1}}{c^2} \left[
        a^{\mu} + \frac{c^2}{\rho h} \Delta^{\mu \nu} \nabla_{\nu} p 
    \right]
    \nonumber \\
    &+
    \frac{ \rho \kappa T^2}{m_b (e + p) c^2} 
    \Delta^{\mu\alpha} \nabla_{\alpha} \left( \frac{\mu}{T}\right)
    \, , \\
\label{eq:T-mu-nu-def-general-param}
    \mathcal{T}_{\mu \nu} 
    \equiv& 
    -
    2\eta\, \sigma_{\mu \nu}
    \,, \\
\label{eq:N-def-BDNK}
    \mathcal{N} 
    \equiv& 
    \rho
    \,, \\
\label{eq:diffusion-current-BDNK-def}
    \mathcal{J}^{\mu} 
    \equiv&
    0\,,
\end{align}
\end{subequations}
where $T$ denotes the temperature, 
$\mu$ the chemical potential, $m_b$ the baryon mass, 
$\zeta$ the bulk viscosity, $\eta$ the shear (dynamic) viscosity, 
and $\kappa$ the thermal conductivity\footnote{We note that BDN instead made use
of the heat coefficient $\sigma$, which is related to $\kappa$ via~\cite{Kovtun_2019}  
\begin{align}
    \sigma
    \equiv 
    \frac{n^2 T}{\left(e+p\right)^2}\kappa
    .
\end{align}}.
The additional transport coefficients $\left(\tau_{\epsilon,i}, \tau_{p,i} \right)$ encode deviations from equilibrium, and will play a vital role in ensuring the
equations of motion are causal and hyperbolic.
The form of $\mathcal{Q}^{\mu}$ is motivated by the fact that in thermal equilibrium,
the heat vector should vanish and the equations of motion should reduce to those
of a perfect fluid (for more discussion see \cite{Bemfica:2020zjp}).
The expansion $\theta$ and the shear tensor $\sigma_{\mu \nu}$ are defined by
\begin{align}
\label{eq:def_fluid_expansion}
\theta 
    &\equiv
    \nabla_{\mu}u^{\mu} 
    ,\\
\label{eq:def_fluid_shear}
\sigma^{\mu\nu} 
    &\equiv 
    \frac{1}{2} \left(
        \Delta^{\mu \gamma} \Delta^{\nu \delta} \nabla_{\gamma} u_{\delta} +
        \Delta^{\mu \gamma} \Delta^{\nu \delta}  \nabla_{\delta} u_{\gamma} 
        \right. \nonumber\\
        &\hspace{1cm}
        \left.-
        \frac{2}{3} \Delta^{\mu \nu}\Delta^{\gamma \delta} \nabla_{\gamma} u_{\delta}
    \right) \,.
\end{align}
For such a BDNK fluid, a choice of the functional dependence of $(\tau_{\epsilon,i}, \tau_{p,i},\tau_{Q,1},\zeta,\eta,\sigma)$ with respect to other equilibrium functions (such as $\epsilon$ and $\rho$) defines a ``hydrodynamic frame.''

A more restrictive first-order fluid model was studied by
BDN~\cite{Bemfica:2020zjp}.
The transport coefficients in our more general parameterization,
Eq.~\eqref{eq:parameterization-general}, 
are related to those in the model studied by BDN~\cite{Bemfica:2020zjp} by the following
\begin{subequations}
\label{eq:general_to_bdn}
\begin{align}
   \tau_{\epsilon,1} &= \rho \tau_{\epsilon}\,,  \\
    \tau_{\epsilon,2} &= p \tau_{\epsilon}\,, \\
    \tau_{p,1} &= \tau_{p} \,, \\ 
    \tau_{p,2} &= p \tau_{p}\,,\\
    \tau_{Q,1} &= \tau_Q \rho h\,,
\end{align} 
\end{subequations}
where $h$ is the specific enthalpy (e.g. \cite{Rezzolla-Book}) 
\begin{align}
    h
    \equiv
    \frac{e+p}{\rho}
    =
    c^2 
    +
    \epsilon 
    +
    \frac{p}{\rho}
    .
\end{align}
and the rest of the transport coefficients ($\tau_{\epsilon,i}, \tau_{p,i}$) are zero.
To derive the relations in Eq.~\eqref{eq:general_to_bdn}, we made use of the continuity
equation $\nabla_{\mu}J^{\mu}=0$, which implies $u^{\mu}\nabla_{\mu}\rho + \rho\theta=0$,
to set $u^{\alpha}\nabla_{\alpha}\rho=-\rho\theta$.
We emphasize that we will make use of the parameterization 
Eq.~\eqref{eq:parameterization-general} throughout this article.

The field equations are given by
\begin{subequations}\label{eq:field-equations}
    \begin{align}
        \label{eq:t-mu-nu-conservation}
        \nabla_{\mu} T^{\mu \nu}&=0\,,\\
        \label{eq:j-mu-conservation}
        \nabla_{\mu} J^{\mu} &=0\,.
\end{align}
\end{subequations}
We project the stress-energy conservation equation [Eq.~\eqref{eq:t-mu-nu-conservation}] 
into components parallel and perpendicular to $u^{\mu}$ to get
\begin{subequations}\label{eq:fluid-equations-projected}
    \begin{align}
\label{eq:energy_equation_general}
        &D\mathcal{E}
        +
        \left(\mathcal{E} + \frac{1}{c^2}\mathcal{P}\right)\theta
        +
        \theta_Q
        \nonumber\\
        &+
        \frac{1}{c^2}\left(
            \mathcal{Q}_{\nu}a^{\nu}
            -
            2\eta\sigma_{\alpha \beta}\sigma^{\alpha \beta}
        \right)
        =
        0
        \,,\\
\label{eq:momentum_equation_general}     
        &a^{\mu} (\mathcal{E} + \frac{1}{c^2}\mathcal{P}) 
         + 
         \Delta^{\mu \alpha}\nabla_{\alpha} \mathcal{P}  
         + 
         \Delta^{\mu}_{\nu} D Q^{\nu} 
                  \nonumber\\
         &- 
         \Delta^{\mu}_{\beta}\nabla_{\alpha}\left( 2\eta \sigma^{\alpha \beta} \right) 
         \nonumber\\
         &
         +
         Q^{\alpha} \left( \omega^{\mu}{}_{\alpha} + \sigma_{\alpha}^{\mu} + \frac{4}{3} \theta \Delta_{\alpha} ^{\mu} \right) 
         =
         0
         \,,
\end{align}
\end{subequations}
where Eq.~\eqref{eq:energy_equation_general} is the projection parallel to $u^{\mu}$ and 
Eq.~\eqref{eq:momentum_equation_general} is the projection perpendicular to $u^{\mu}$. 
We call Eq.~\eqref{eq:energy_equation_general} the ``energy equation'' and
Eq.~\eqref{eq:momentum_equation_general} the ``momentum equation''.
The conservation of the rest mass current $\nabla_{\mu} J^{\mu} = 0$ with our parameterization [Eq.~\eqref{eq:N-def-BDNK}-\eqref{eq:diffusion-current-BDNK-def}] can be simplified to
\begin{align}
    D \rho + \rho \theta &=0\,.
\end{align}
In the above equations, we have also introduced the convective
derivative $D$, the heat expansion $\theta_Q$,
and the vorticity $\omega_{\mu\nu}$ via
\begin{subequations}
\begin{align}
    D
    &\equiv
    u^{\alpha}\nabla_{\alpha}
    ,\\
    \theta_Q 
    &\equiv 
    \nabla_{\mu} \mathcal{Q}^{\mu}\,, \\
    \omega_{\mu \nu} 
    &\equiv 
    \Delta_{\mu}^{\gamma}\Delta_{\nu}^{\delta}\nabla_{[{\delta}}u_{\gamma]}
    \,.
\end{align} 
\end{subequations}
\subsection{Causality Constraints}\label{subsec:Causality-BDNK-general}
We next write down the constraints the transport coefficients 
must satisfy in order for the relativistic equations of motion to be hyperbolic and causal.
We encountered analogous constraints for the transport coefficients of a toy model in 
Sec.~\ref{sec:toy-models} (see, Eq.~\eqref{eq:causality-toy-model}).

To simplify the algebraic form of the causality constraints, 
we introduce the following definitions
\begin{subequations}\label{eq:new-definitions-transport-coefficients}
  \begin{align}
    \varphi 
    &\equiv 
    \frac{\mu}{T} 
    \,,\\ 
    \partial_{\epsilon} 
    &\equiv
    \left.\frac{\partial}{\partial{\epsilon}}\right|_{\rho} 
    \,,  
    \partial_{\rho} 
    \equiv
    \left.\frac{\partial}{\partial{\rho}}\right|_{\epsilon}
    \,, \\
    \tau_{\epsilon,\parallel} 
    &\equiv 
    \tau_{\epsilon,1} 
    + 
    \frac{\tau_{\epsilon,3} \partial_{\epsilon} \varphi}{c^2} 
    \,,\\
    \tau_{\epsilon,\perp} 
    &\equiv  
    \frac{c^2 \tau_{Q,1}}{\rho h } \partial_{\epsilon} p 
    + 
    \frac{n \kappa T^2}{\rho h} \partial_{\epsilon} \varphi 
    \,, \\
    \tau_{\rho,\perp} 
    &\equiv 
    \frac{c^2 \tau_{Q,1}}{\rho h } \partial_{\rho} p 
    + 
    \frac{n \kappa T^2}{\rho h} \partial_{\rho} \varphi  
    \,,\\
    \tau_{\epsilon v} 
    &\equiv 
    \tau_{\epsilon,\perp} 
    + 
    c^2\tau_{p,1} 
    + 
    \tau_{p,3} \partial_{\epsilon} \varphi 
    \,,\\ 
    V_1 
    &\equiv 
    \tau_{p,2} 
    -
    \zeta 
    - 
    \frac{4 \eta}{3} 
    - 
    \frac{\rho \tau_{\rho,\perp}}{c^2}
    - 
    \frac{\rho \tau_{p,3} \partial_{\rho} \varphi}{c^2} 
    \,,\\
    \tau_{u} 
    &\equiv 
    \tau_{Q,1} 
    + 
    \tau_{\epsilon,2} 
    - 
    \frac{\rho \tau_{\epsilon,3} \partial_{\rho} \varphi}{c^2}
    \,.
\end{align}  
\end{subequations}
Note that $\varphi$ is sometimes called the fugacity, 
and is constant in thermal equilibrium \cite{Rezzolla-Book}.

With the above definitions, we can now derive the conditions required for the field equations [Eq.~\eqref{eq:field-equations}] to remain causal and hyperbolic. As we explain in Appendix~\ref{appendix:principal-symbol-general}, one can do so by calculating the principal symbol of the equations of motion and requiring that the characteristic speeds be real and smaller than the speed of light. Doing so, we find the following causality constraints  
\begin{subequations}\label{eq:causality-general-model}
\begin{align}
\label{eq:causality-constraints-1-general}
    \tau_{Q,1} > \eta\,, \\
\label{eq:causality-constraints-2-general}
    a_2^2 \geq 4 a_1 a_3 >0\,, \\
\label{eq:causality-constraints-3-general}
    2 a_1 > a_2 > 0\,,\\
\label{eq:causality-constraints-4-general}
    a_1 + a_3 > a_2 > 0\,,
\end{align}
\end{subequations}
where
\begin{subequations}\label{eq:ai-defs}
   \begin{align}
    a_1 &= c^2\tau_{Q,1} \tau_{\epsilon,\parallel}\,, \\
    a_2 &= \tau_{\epsilon v} \tau_{u} - c^2V_1\tau_{\epsilon,\parallel} - \tau_{\epsilon,\perp} \tau_{Q,1}\,,  \\
    \label{eq:a3-def}
    a_3 &= V_1 \tau_{\epsilon,\perp} + \frac{\rho \tau_{\epsilon v} \tau_{\rho,\perp}}{c^2}\,.
\end{align} 
\end{subequations}
The above constraints generalize the analysis of 
BDN \cite{Bemfica:2020zjp}, which focused on the simplified BDNK fluid model of Eq.~\eqref{eq:general_to_bdn} (i.e.,~the above equations are the hyperbolicity constraints for the general BDNK fluid model of Eq.~\eqref{eq:parameterization-general}, which is the most general
extension of an Eckart-like fluid).
The causality constraints can be interpreted as restricting 
the physically allowed values of the transport coefficients $\tau_{\epsilon,i},\tau_{p,i}$ 
for BDNK fluids.
If one imposes linear mode stability of the equations of motion about thermal equilibrium,
the transport coefficients must satisfy a further set of constraint equations~\cite{Bemfica:2020zjp}.
These can be found using the same method as in~\cite{Bemfica:2020zjp} after using the definitions introduced in Eq.~\eqref{eq:new-definitions-transport-coefficients}.
\section{Non-relativistic limit of the BDNK fluid model}\label{sec:NR-limit-BDNK}
In this section, we derive the non-relativistic limit of the conservation equations
~\eqref{eq:field-equations} for a BDNK fluid, neglecting the effects of gravity 
(i.e., we consider the non-relativistic limit about Minkowski spacetime).
We first show in Sec.~\ref{subsec:NR-limit-general}
that the non-relativistic limit of the BDNK equations of motion are never hyperbolic.
Next, we determine the conditions under which the non-relativistic 
equations of motion reduce to the Navier-Stokes equations.
We show that this reduction and the fact that relativistic motion is causal imposes important restrictions on the transport coefficients appearing in the non-relativistic limit.
These constraints can be used to 
derive an upper bound on the ratio of the shear viscosity to the entropy density,
which we show in Sec.~\ref{subsec:NR-limit-NS-condition}.
Finally, we discuss some implications of our results in Sec.~\ref{subsec:Discussion}.

In deriving the non-relativistic limit, we make the following assumptions:
\begin{enumerate}[leftmargin=*]
\setlength\parskip{0.0cm}
\setlength\parsep{0.0cm}
\item \textbf{Non-zero non-relativistic viscosity:} The non-relativistic limit of the hydrodynamical 
equations of motion of a BDNK fluid must also be those of a viscous fluid.
This is equivalent to the requirement that the non-relativistic limit of at least one of 
the coefficients $\eta, \zeta$ and $ \kappa$ is nonzero.
\item \textbf{Preservation of the structure of the post-Newtonian metric}: 
We assume that the stress-energy tensor component $c^{-1} T^{0j}$ has the following non-relativistic expansion
\begin{align}\label{eq:T0j-assumption}
    c^{-1} T^{0j} &= T_0^j  + T_2^j c^{-2} + T_4^j c^{-4} + \mathcal{O}(c^{-5})\,.
\end{align}
This assumption is used to ensure that the structure of the post-Newtonian metric remains the same as that of a perfect fluid~\cite{Poisson-Will}.
\end{enumerate}
We note that calculation of transport coefficients from relativistic kinetic theory respects the first assumption and also predicts the same expansion for the stress-energy
tensor \cite{Israel-1963,Stewart-1971,cercignani2002relativistic}.
What is the significance of Assumption 2? 
This requirement essentially requires that relativistic corrections to the spacetime metric 
(that is, corrections that come in with an inverse power of $c$) should take
the same general form as they do for perfect fluid matter. In particular, this
condition enforces that the metric does not receive post-Newtonian corrections that scale
with odd powers of $c$ until 2.5 post-Newtonian order\cite{Poisson-Will}.

In this section, we will not restrict our analysis to any specific hydrodynamic frame, but let us point out that the frame used in~\cite{Pandya:2022sff} 
does not satisfy the two assumptions presented above. 
The model studied in~\cite{Pandya:2022sff} used the simple parameterization by BDN [Eq.~\eqref{eq:general_to_bdn}].
In their hydrodynamic frame, $(\tau_Q,\tau_{\epsilon},\tau_p) = \mathcal{O}(c^{-1}) = (\eta,\zeta,\kappa)$. 
From this scaling, one can show that $(1/c)T^{0i}$ does not scale with even powers
of $1/c^2$, as required by Eq.~\eqref{eq:T0j-assumption}.
Moreover, we see that this frame violates the assumption of non-zero non-relativistic viscosity.
One can verify that in the non-relativistic limit, the equations of \cite{Pandya:2022sff} reduce to the Euler-equations.
Although the hydrodynamic frame chosen in  \cite{Pandya:2022sff} is perfectly valid
from a mathematical point of view (the equations of motion can be shown to satisfy
the BDNK causality constraints \cite{Pandya:2022sff}), 
it cannot be used to model relativistic fluid motion of astrophysical interest.
More general hydrodynamic frames need to be derived for the 
BDNK model to be able to describe more realistic fluid flows.

We recall the Navier-Stokes equations without external forces~\cite{1961hhs..book.....C} 
\begin{subequations}\label{eq:Navier-Stokes-equations.}
  \begin{align}
    \label{eq:conservation-dens-NRNS}
        &D^N \rho + \rho \theta^N 
        =0
        \,,\\
    \label{eq:conservation-internal-energy-NRNS}
        &\rho D^N \epsilon + p \theta^N + \partial_j Q^j 
        \nonumber\\
        &
        - 
        \zeta \left( \theta^N \right)^2 
        - 
        2\eta \left(\sigma^N\right)^{ij}\sigma^N_{ij}
        =
        0
        \,,\\
\label{eq:momentum-equation-NRNS}
    &\rho D^N v^j 
    + 
    \partial^j p 
    -
    \partial_k \left[
        \zeta \theta^N \delta^{jk} 
        + 
        2 \eta \left(\sigma^N\right)^{jk}
    \right]
    =0
    \,.
\end{align}  
\end{subequations}
where
\begin{align}
    D^N &\equiv \partial_t + v^j \partial_j \,, \\
    \sigma^N_{ij} &\equiv \partial_{(i}v_{j)} - \frac{1}{3}\delta_{ij}\theta^N\,,\\
    \theta^N &= \partial_i v^i
    .
\end{align}
The subscript N stands for “Newtonian” and we use it to distinguish between 
the fully relativistic quantities used in Eq.~\eqref{eq:fluid-equations-projected}.
Equation~\eqref{eq:conservation-dens-NRNS} expresses the conservation of mass, 
Eq.~\eqref{eq:conservation-internal-energy-NRNS} describes the evolution of internal energy and
Eq.~\eqref{eq:momentum-equation-NRNS} describes the evolution of the 3-velocity of the fluid.
We note that Eqs.~\eqref{eq:conservation-dens-NRNS}-\eqref{eq:momentum-equation-NRNS} are not closed, 
as we have not prescribed how the heat $Q^i$, the viscosity coefficients $\eta,\zeta$, or the
pressure $p$ depend on $\rho$ and $\epsilon$.
A common choice for the heat flux is the Fourier law~\cite{1961hhs..book.....C}
\begin{align}\label{eq:Fourier-law}
    Q^j_{NSF} &= -\kappa \partial^j T\,,
\end{align}
and the resulting equations are then called the Navier-Stokes-Fourier equations~\cite{1961hhs..book.....C}.
\subsection{Character of the non-relativistic equations}\label{subsec:NR-limit-general}
To derive the non-relativistic limit of the equations of motion,
we pick a local Lorentz frame $x^{\alpha} = (ct, x, y,z)$.
The line element is then given by
\begin{equation}
    -c^2 d\tau^2 
    = 
    \eta_{\alpha \beta} dx^{\alpha} dx^{\beta} 
    ,
\end{equation}
where $\eta_{\alpha \beta}$ is the Minkowski metric,
$\eta_{\alpha \beta} = \text{diag} (-1,1,1,1)$.
The four-velocity of the fluid is defined by
\begin{equation}
    u^{\alpha} \equiv \frac{d z^{\alpha}}{d\tau} \,,
\end{equation}
where $z^{\alpha}$ is the four-position vector of a fluid element. 
The local Lorentz factor $W$ is defined by
\begin{equation}
    W \equiv \frac{d t}{d\tau}\,.
\end{equation}
The normalization of the 4-velocity $u^a u_a = -c^2$ implies that 
$u^a = \left(cW, W v^i \right)$, where
\begin{equation}\label{eq:lorentz-factor}
     W 
    \equiv 
    \frac{1}{\sqrt{1 - c^{-2}v_iv^i}}\,.
\end{equation}
We will also sometimes use the following shorthands
\begin{subequations}
\label{eq:shorthands_E_Ej}
\begin{align}
    E &\equiv \rho D^N \epsilon + p \theta^N \,,\\
    E^j &\equiv \rho D^N v^j + \partial^j p \,.
\end{align}
\end{subequations}

We begin by studying some simple consequences of the assumptions
of non-zero non-relativistic viscosity and the preservation of the structure of the post-Newtonian metric.
From Eq.~\eqref{eq:decomp_tmunu}, we see that $c^{-1} T^{0j}$ is given by
\begin{align}
    \label{eq:expansion_T0j}
    \frac{1}{c} T^{0j} 
    =& 
    \left( \mathcal{E} + \frac{\mathcal{P}}{c^2} \right) W^2 v^j 
    \nonumber\\
    &+
    \mathcal{Q}^j W + \frac{\mathcal{Q}^0 v^j W}{c} 
    - 
    2 \eta \frac{\sigma^{0j}}{c}
    \,.
\end{align}
From the form of the Lorentz factor in Eq.~\eqref{eq:lorentz-factor}, we know that $W$ only has even powers of $c^{-1}$ 
in its non-relativistic expansion.
Similarly, one can calculate the non-relativistic expansion of the time-space components of the shear tensor and conclude that $\sigma^{0j}$ contains only odd powers of $c^{-1}$
(for more detailed calculations see Appendix~\ref{appendix:details-nr-expansion}).
Therefore, to satisfy the preservation of the structure of the post-Newtonian metric, 
we see that we must have the following non-relativistic expansions
\begin{subequations}
\begin{align}
    \mathcal{E} &= \mathcal{E}_0 + c^{-2} \mathcal{E}_{2} + c^{-4} \mathcal{E}_{4} + \mathcal{O}(c^{-5})\,, \\
    \mathcal{Q}^j &= \mathcal{Q}^j_0 + c^{-2} \mathcal{Q}^j_{2} + c^{-4} \mathcal{Q}^j_{4} + \mathcal{O}(c^{-5})\,, \\
    \mathcal{Q}^0 &= c^{-1} \mathcal{Q}^0_1 + c^{-3} \mathcal{Q}^0_{3} + \mathcal{O}(c^{-5})\,, \\
    \mathcal{P} &= \mathcal{P}_0 + c^{-2} \mathcal{P}_{2} + \mathcal{O}(c^{-3})\,.
\end{align}
\end{subequations}
From $\mathcal{Q}^{\mu} u_{\mu} = 0$, we see that, if $\mathcal{Q}^j = \mathcal{Q}^j_0 + \mathcal{O}(c^{-2})$, then $\mathcal{Q}^0_1 = v_j Q^j_0$.
We also assert that $\mathcal{E}_0 = \rho$;
this assumption reflects the fact that for non-relativistic motion, the largest contribution to the energy comes from the rest-mass energy.
We note though that while the results we derive below do not depend on this assumption,
it makes the presentation and the physical intuition clearer.

By a straightforward calculation, one can show that the above equations, along
with our assumption of non-zero non-relativistic viscosity, imply 
that the BDNK transport coefficients must have the following non-relativistic expansion
\begin{subequations}\label{eq:nr-expansion-transport-coeffs}
    \begin{align}
        \tau_{\epsilon,i} &= 
        \tau_{\epsilon,i}^{(0)} + 
        \frac{\tau_{\epsilon,i}^{(2)}}{c^2} 
        + \mathcal{O}(c^{-4}) \,,\\
        \tau_{p,i} &= \tau_{p,i}^{(0)} + 
        \frac{\tau_{p,i}^{(2)}}{c^2} 
        + \mathcal{O}(c^{-4}) \,,\\
        \tau_{Q,1} &= \tau_{Q,1}^{(-2)} c^2 + \tau_{Q,1}^{(0)} +
        \frac{\tau_{Q,1}^{(2)}}{c^2} 
        + \mathcal{O}(c^{-4}) \,,\\
        \label{eq:eta-expansion-v1}
        \eta &= \eta^{(0)} + c^{-2}\eta^{(2)} + \mathcal{O}(c^{-3})\,, \\
        \zeta &= \zeta^{(0)} + c^{-2} \zeta^{(2)} + \mathcal{O}(c^{-3})\,, \\
    \label{eq:kappa-expansion-v1}
        \kappa &= \kappa^{(0)} + c^{-2}\kappa^{(2)} + \mathcal{O}(c^{-3})\,.
    \end{align}
\end{subequations}
For more details for our expansion of $\mathcal{E},\mathcal{P}$, and $\mathcal{Q}^{\mu}$, 
see Appendix~\ref{appendix:details-nr-expansion}.
We note that relativistic kinetic theory also predicts a similar form for 
$\eta, \zeta$, and $\kappa$, namely that they are even functions of $c$
\cite{Israel-1963,Stewart-1971,cercignani2002relativistic}.

We next outline why $\tau_{Q,1}^{(-2)}=0$.
The leading order term in the expansion in $(1/c)$ of the energy equation 
[Eq.~\eqref{eq:energy_equation_general}] 
is (see Eq.~\eqref{eq:PN-Minkowski-expansion-energy-equation})
\begin{align}
    \partial_j \mathcal{Q}^j_0 + \mathcal{O}(c^{-2}) 
    &= 
    0 \nonumber\\
    \implies
    \partial_j \left(\frac{\tau_{Q,1}^{(-2)}}{\rho} E_j \right)+ \mathcal{O}(c^{-2}) 
    &= 
    0\,.
\end{align}
To obtain the second line, we used Eq.~\eqref{eq:Q-j-PN-Mink-2-def}
to expand $Q^j$ to leading order in $1/c$.
In order to obtain the correct non-relativistic equations of motion for the energy
equation, we see that we need $\mathcal{Q}_0^j=0$, which implies $\tau_{Q,1}^{(-2)}=0$.

With this, we write down the non-relativistic limit of 
the conservation law in Eq.~\eqref{eq:j-mu-conservation}, 
the energy equation~\eqref{eq:energy_equation_general}, 
and the momentum equation \eqref{eq:momentum_equation_general}:
\begin{widetext}
\begin{subequations}\label{eq:NR-limit-eqs-general}
    \begin{align}
    \label{eq:dens-cons-general}
        &D^N \rho + \rho \theta^N = \mathcal{O}(c^{-2}) \,, \\
    \label{eq:int-energy-cons-general}
        &\rho D^N \epsilon 
        + 
        p \theta^N 
        + 
        \partial_j Q^j_{NSF}  
        + 
        \left[\tau_{\epsilon,2}^{(0)} 
        + 
        \tau_{p,2}^{(0)} 
        - 
        \zeta^{(0)} 
        - 
        \rho \tau_{\epsilon,3}^{(0)} \partial_{\rho} \left(\frac{m_b}{T} \right)
        - 
        \rho \tau_{p,3}^{(0)} \partial_{\rho}\left( \frac{m_b}{T}\right)\right] 
        \left( \theta^N \right)^2 - 2\eta 
        \left(\sigma^N\right)^{ij}\sigma^N_{ij}  
        =
        \nonumber \\ 
        &-
        \partial_j
        \left[ 
        \frac{\tau_{Q,1}^{(0)}}{\rho}
        E^j
        \right]
        -D^N \left[ \left(\tau_{\epsilon,1}^{(0)} + \tau_{\epsilon,3}^{(0)} \partial_{\epsilon} \left(\frac{m_b}{T} \right) \right)D^N \epsilon 
        +
        \left(
        \tau_{\epsilon,2}^{(0)}   - \rho \tau_{\epsilon,3}^{(0)} \partial_{\rho} \left(\frac{m_b}{T} \right)
        \right)
        \theta^N
        \right]
        \nonumber \\
        &- 
        \left(\tau_{\epsilon,1}^{(0)} + \tau_{\epsilon,3}^{(0)} \partial_{\epsilon} \left(\frac{m_b}{T} \right) 
        +
        \tau_{p,1}^{(0)}
        +
        \tau_{p,3}^{(0)} \partial_{\epsilon} \left(\frac{m_b}{T} \right) 
        \right)
        \left(D^N \epsilon\right) \theta^N
        +
        \mathcal{O}(c^{-2})
        \,,\\
    \label{eq:momentum-cons-general}
    &\rho D^N v^j 
    + 
    \partial^j p 
    -
    \partial_k \left[
        \left( 
            \zeta^{(0)} 
            -
            \tau_{p,2}^{(0)} 
            + 
            \rho \tau_{p,3}^{(0)} \partial_{\rho}\left( \frac{m_b}{T}\right)
        \right) 
        \theta^N \delta^{jk} 
        + 
        2 \eta^{(0)} \left(\sigma^N\right)^{kj}
     \right]
     \nonumber \\
    &
    =
    -\partial^j \left[ 
    \left( \tau_{p,1}^{(0)} +
    \tau_{p,3}^{(0)}\partial_{\epsilon}\left( \frac{m_b}{T}\right)
    \right)
    D^N\epsilon
    \right]\,,
    \end{align}
\end{subequations}
\end{widetext}
Several identities for establishing these equations are provided in 
Appendix~\ref{appendix:details-nr-expansion}.
The left-hand side of Eq.~\eqref{eq:NR-limit-eqs-general}
are the Navier-Stokes-Fourier equations~\eqref{eq:Navier-Stokes-equations.} and \eqref{eq:Fourier-law}.
The right-hand side contains additional derivatives arising from the non-relativistic expansion of
the BDNK equations of motion.

Comparing to the Navier-Stokes equations, we see that in order to identify $\zeta^{(0)}$
with the fluid bulk viscosity, and for the fluid equations of motion to only incur changes
at higher order in gradients, we need
\begin{subequations}
\label{eq:zeta-identification}
\begin{align}
    \zeta^{(0)}
    =&
    \zeta^{(0)}
    +
    \rho \tau_{p,3}^{(0)} \partial_{\rho} \left(
    \frac{m_b}{T}\right)
    -
    \tau_{p,2}^{(0)}
    \nonumber\\
    &
    +
    \rho \tau_{\epsilon,3}^{(0)}
    \partial_{\rho} \left(
    \frac{m_b}{T}\right)
    -
    \tau_{\epsilon,2}^{(0)}
    ,\\
    \zeta^{(0)} =& 
    \zeta^{(0)} 
    + 
    \rho \tau_{p,3}^{(0)} \partial_{\rho}\left( \frac{m_b}{T}\right)
    -
    \tau_{p,2}^{(0)} 
    .
\end{align}
\end{subequations}
The first line comes from comparing the left-hand side of 
Eq.~\eqref{eq:int-energy-cons-general} with Eq.~\eqref{eq:conservation-internal-energy-NRNS}
and the second line comes from comparing the left-hand side of 
Eq.~\eqref{eq:momentum-cons-general} with Eq.~\eqref{eq:momentum-equation-NRNS}. 
From these two relations, we conclude that
\begin{subequations}
\begin{align}
    \label{eq:condition-zeta}
    \rho \tau_{p,3}^{(0)} \partial_{\rho} \left(
    \frac{m_b}{T}\right)
    -
    \tau_{p,2}^{(0)}
    &=
    0
    ,\\
    \rho \tau_{\epsilon,3}^{(0)}
    \partial_{\rho} \left(
    \frac{m_b}{T}\right)
    -
    \tau_{\epsilon,2}^{(0)} 
    &=
    0
    .
\end{align}
\end{subequations}
These relations ensure that $\zeta^{(0)}$ can be interpreted as the fluid bulk viscosity, and they will simplify our future equations.

To analyze the character of Eq.~\eqref{eq:NR-limit-eqs-general},
we consider a linear analysis about a homogeneous fluid state with zero velocity.
We show that Eq.~\eqref{eq:NR-limit-eqs-general} has indefinite character 
as a set of partial differential equations (i.e.,~the equations are not purely hyperbolic)
even about a flat background spacetime, 
which is enough to conclude that the non-relativistic equations of motion are not hyperbolic. We first perturb Eq.~\eqref{eq:NR-limit-eqs-general} with
\begin{equation}
\begin{aligned}
    \epsilon \to& \epsilon + \delta \epsilon\left(t,x^i\right)
    \,,\\
    \rho \to& \rho + \delta \rho\left(t,x^i\right) 
    \,,\\
    v^j \to& \delta v^j\left(t,x^i\right)  
    \,,\\
\end{aligned}
\end{equation}
and obtain
\begin{subequations}\label{eq:perturbed-equation-general}
\begin{align}
    &\partial_t \delta \rho + \rho \partial_j \delta v^j = 0 \,, \\
    &\left(
        \tau_{\epsilon,\parallel}^{(0)} \partial_t^2 
        +
        \tau_{\epsilon,\perp}^{(0)} \Delta
        +
        \rho
        \partial_t
    \right)
    \delta \epsilon
    \nonumber \\
    &
    +
    \tau_{\rho,\perp}^{(0)} \Delta \delta \rho
    +
    \left(\tau_{u}^{(0)}  \partial_t + p \right) \partial_j \delta v^j
    =0 
    \,,\\
    &
    \left(
        \tau_{\epsilon v}^{(-2)} \partial_t
        +
        \partial_{\epsilon}p
    \right)
    \partial^j \delta \epsilon
    +
    \hat{C}^{j m n}_{l} \partial^2_{m n}
    \delta v^l
    \nonumber\\
    &
    +
    \rho\partial_t\delta v^j 
    +
    \left(\partial_{\rho}p\right)\partial^j\delta\rho
    =0
    \,.
\end{align}
\end{subequations}
where $\Delta = \partial_j \partial^j$ is the flat space Laplacian.
In the above equations, 
we have used the definitions of the transport coefficients given in 
Eq.~\eqref{eq:new-definitions-transport-coefficients}, which in the
non-relativistic limit are
\begin{subequations}
    \begin{align}
        \tau_{\epsilon,\parallel}^{(0)} &= \tau_{\epsilon,1}^{(0)} + \tau_{\epsilon,3}^{(0)} \partial_{\epsilon} \left( \frac{m_b}{T}\right) \, ,\\
        \tau_{\epsilon,\perp}^{(0)} &= \frac{\tau_{Q,1}^{(0)}}{\rho} \partial_{\epsilon} p 
        +
        \frac{n \kappa^{(0)} T^2}{\rho}\partial_{\epsilon}\left( \frac{m_b}{T}\right) \,,\\
        \tau_{\rho,\perp}^{(0)} &= \frac{\tau_{Q,1}^{(0)}}{\rho} \partial_{\rho} p 
        +
        \frac{n \kappa^{(0)} T^2}{\rho}\partial_{\rho}\left( \frac{m_b}{T}\right) \,,\\
        \tau_{u}^{(0)} &= \tau_{Q,1}^{(0)} + \tau_{\epsilon,2}^{(0)} - \rho \tau_{\epsilon,3}^{(0)} \partial_{\rho} \left( \frac{m_b}{T}\right) 
        =\tau_{Q,1}^{(0)} \,,\\
        \tau_{\epsilon v}^{(-2)} &= \tau_{p,1}^{(0)} + \tau_{p,3}^{(0)} \partial_{\epsilon} \left( \frac{m_b}{T}\right) \,,\\
        \hat{C}^{j m n}_{l} &= -\left(\zeta^{(0)} +\frac{\eta^{(0)}}{3} \right)\delta^{j(m}\delta^{n)}_l - \eta^{(0)} \delta^{mn}\delta^j_l\,,
\end{align}
\end{subequations}
we have used the identity given in Eq.~\eqref{eq:condition-zeta} in the fourth line and used the definition of $\zeta^{(0)}$ [Eq.~\eqref{eq:zeta-identification}] in the final line.
One can analyze the causal properties of Eq.~\eqref{eq:perturbed-equation-general} 
by studying the principal symbol of this system \cite{kreiss_oliger_ns_eqns_book}. 
This analysis is carried out in Appendix~\ref{appendix:causality-general-NR} and we find that the system is non-hyperbolic.
For example, there is a diffusive shear mode with dispersion relation
\begin{align}
    i\omega = -\frac{\eta^{(0)}}{\rho} k^2\,.
\end{align}
To summarize, the non-relativistic limit of the BDNK fluid model is not hyperbolic. 
\subsection{Non-relativistic limit and reduction to the Navier-Stokes equation}\label{subsec:NR-limit-NS-condition}
Here we determine the conditions under which Eq.~\eqref{eq:NR-limit-eqs-general} 
reduces \emph{exactly} to the Navier-Stokes system of Eq.~\eqref{eq:Navier-Stokes-equations.},
and study how this reduction affects the constraints coming from causality for the relativistic equations.
We see that when 
\begin{subequations}\label{eq:NS-expansion-relations}
    \begin{align}
        \tau_{\epsilon,\parallel}^{(0)} 
        = 
        \tau_{\epsilon,1}^{(0)} 
        + 
        \tau_{\epsilon,3}^{(0)} \partial_{\epsilon} \left( \frac{m_b}{T}\right) 
        &= 
        0 
        \,,\\
        \tau_{\epsilon v}^{(-2)} 
        = 
        \tau_{p,1}^{(0)} 
        + 
        \tau_{p,3}^{(0)} \partial_{\epsilon} \left( \frac{m_b}{T}\right) 
        &= 0\,,
    \end{align}
\end{subequations}
then Eqs.~\eqref{eq:NR-limit-eqs-general} reduce to
\begin{subequations}\label{eq:NR-limit-eqs-general-with-expansion-relations} 
\begin{align}
    &
    D^N \rho 
    + \rho \theta^N 
    = 
    \mathcal{O}\left(c^{-2}\right)
    \,,\\
    &
    \rho D^N \epsilon 
    + 
    p \theta^N 
    + 
    \partial_j \mathcal{Q}^j_{0} 
    - 
    \zeta \left( \theta^N \right)^2 
   \nonumber\\ 
    &
    - 
    2\eta \sigma_{ij}^N\left(\sigma^N\right)^{ij}
    =
    \mathcal{O}\left(c^{-2}\right)
    \,,\\
    &
    \rho D^N v^j + \partial^j p 
    \nonumber\\
    &
    -
    \partial_k \left[
        \zeta \theta^N \delta^{jk} 
        + 
        2 \eta \left(\sigma^N\right)^{jk}
    \right]
    =
    \mathcal{O}(c^{-2})
    \,,
\end{align}  
\end{subequations}  
where
\begin{align}\label{eq:heat-flux-BDNK-NR}
    \mathcal{Q}^j_{0} 
    =
    - 
    \kappa^{(0)} \partial^j T
    +
    \frac{\tau^{(0)}_{Q,1}}{\rho} E^j 
    \,.
\end{align}
The system of Eq.~\eqref{eq:NR-limit-eqs-general-with-expansion-relations}
takes almost the same form as the Navier-Stokes-Fourier equations 
\eqref{eq:Navier-Stokes-equations.}, 
except that the heat flux of Eq.~\eqref{eq:heat-flux-BDNK-NR} contains the gradient correction
$\tau_{Q,1}^{(0)}E^j/\rho$ (see Eq.~\eqref{eq:shorthands_E_Ej}).
This term is generically non-zero because the causality constraint of Eq.~\eqref{eq:nr-causality_constraint-shear} implies that
$\tau_{Q,1}^{(0)}>0$ as long as $\eta^{(0)}>0$, which we have assumed to be true.
We conclude that the non-relativistic limit of a BDNK fluid 
cannot be consistent with the Fourier-law in Eq.~\eqref{eq:Fourier-law}.
This result holds regardless whether the energy and momentum equations
reduce to the Navier-Stokes equations.

We can use the causality constraints, combined with the conditions of Eq.~\eqref{eq:NS-expansion-relations},
to derive an inequality for the ratio of thermal conductivity to the shear viscosity.
First we expand Eq.~\eqref{eq:causality-constraints-1-general}, and find that
\begin{align}
    \label{eq:nr-causality_constraint-shear}
    \tau_{Q,1}^{(0)} > \eta^{(0)} + \mathcal{O}(c^{-2})
    \,,
\end{align}
which implies that $\tau_{Q,1}^{(0)}$ cannot be zero.
We next expand $a_3>0$ [Eq.~\eqref{eq:ai-defs}] to leading order in $1/c$. 
Using Eq.~\eqref{eq:NS-expansion-relations}, we obtain
\begin{subequations}\label{eq:a_3-geq-NS-simplification}
    \begin{align}
        &a_3 
        = 
        -
        \left( \zeta^{(0)} + \frac{4 \eta^{(0)}}{3} \right)\tau_{\epsilon,\perp}^{(0)} + \mathcal{O}(c^{-2}) > 0 \nonumber \\
        &\implies \tau_{\epsilon,\perp}^{(0)} < \mathcal{O}(c^{-2}) \,,\\
        &\tau_{\epsilon,\perp}^{(0)}  = \frac{\tau_{Q,1}^{(0)}}{\rho} \partial_{\epsilon}p - \kappa^{(0)} \partial_{\epsilon} T + \mathcal{O}(c^{-2}) <\mathcal{O}(c^{-2})\,, \nonumber \\
       &\implies \frac{\tau_{Q,1}^{(0)}}{\rho} <\kappa^{(0)} \frac{ \partial_{\epsilon} T}{\partial_{\epsilon}p}+ \mathcal{O}(c^{-2}) \,.
    \end{align}
\end{subequations}
Combining the above inequality with Eq.~\eqref{eq:nr-causality_constraint-shear} we find
\begin{align}\label{eq:inequality-thermal-conductivity}
    \rho \kappa^{(0)}>\eta^{(0)} \frac{ \partial_{\epsilon} p}{\partial_{\epsilon}T}+ \mathcal{O}(c^{-2})\,.
\end{align}
The most salient feature of this inequality is that it does not involve the 
coefficients $\tau_{\epsilon,i}$ or $\tau_{\rho,i}$.
We can use Eq.~\eqref{eq:inequality-thermal-conductivity} to put an upper bound
on $\eta^{(0)}/s$ (where, $s$ is the entropy density), by making use of the following
thermodynamic relation~\cite{Rezzolla-Book}
\begin{align}
    dp 
    = 
    s d T + n d \mu 
    \implies  
    \frac{\partial_{\epsilon} p}{\partial_{\epsilon} T} 
    = 
    s + n \frac{\partial_{\epsilon} \mu}{ \partial_{\epsilon} T}
    .
\end{align}
Therefore, Eq.~\eqref{eq:inequality-thermal-conductivity} becomes
\begin{align}\label{eq:upper-bound-eta-by-s}
    \frac{
        \rho \kappa^{(0)}
    }{
        s \left(s + n \frac{\partial_{\epsilon} \mu}{ \partial_{\epsilon} T} \right)
    }
    > 
    \frac{\eta^{(0)}}{s} 
    + 
    \mathcal{O}(c^{-2})
    \,.
\end{align}

To summarize, 
the non-relativistic limit of a BDNK fluid model reduces \emph{exactly}
to the Navier-Stokes equation if Eq.~\eqref{eq:NS-expansion-relations} is satisfied. 
Nevertheless, the Fourier-law [Eq.~\eqref{eq:Fourier-law}] 
always receives corrections by higher-order gradient terms [Eq.~\eqref{eq:heat-flux-BDNK-NR}]
to ensure causality for relativistic motion.
Additionally, the causality constraints, combined with the requirement that
the non-relativistic equations of motion limit exactly to the
Navier-Stokes equations,
imposes a constraint [Eq.~\eqref{eq:inequality-thermal-conductivity}]
on the value of the shear viscosity.

We note that the inequality of Eq.~\eqref{eq:inequality-thermal-conductivity}
can be used in conjunction with the KSS conjecture~\cite{Kovtun:2004de} 
to obtain a lower bound on $\kappa^{(0)}$.
The KSS conjecture states that the ratio $\eta/s$ 
for any fluid must be greater than  
\begin{align}
    \label{eq:kss-bound}
    \frac{\eta^{(0)}}{s} \geq \frac{\hbar}{4 \pi k_B}\,.
\end{align}
One can combine Eq.~\eqref{eq:kss-bound} with Eq.~\eqref{eq:upper-bound-eta-by-s} 
to obtain a lower bound on the thermal conductivity
\begin{align}
    \label{eq:lower-bound-kappa-KSS}
    &\frac{\rho \kappa^{(0)}}{s^2} > \frac{\hbar}{4 \pi k_B} \left(1 + \frac{n}{s} \frac{\partial_{\epsilon} \mu}{ \partial_{\epsilon} T} \right) + \mathcal{O}(c^{-2})\,.
\end{align}
\begin{table*}[htb]
   \caption{Violation of Eq.~\eqref{eq:inequality-ideal-gas} for some common liquids.
   We find that some common viscous liquids violate simplified inequality presented in Eq.~\eqref{eq:inequality-ideal-gas}.
   Common gases generally do not violate Eq.~\eqref{eq:inequality-ideal-gas}.
   We caution the reader that the inequality derived in Eq.~\eqref{eq:inequality-ideal-gas} can only be applied to fluids which retain their fluid properties in both relativistic and non-relativistic regions.
   All of the data below is quoted from~\cite{CRC-Handbook}. }
    \centering
    \begin{tabular}{|c|c|c|c|c|c|c|}
    \hline
        Liquid & $\eta^{(0)}$ ($\mu \text{Pa s}$)  & $\kappa^{(0)}$ (mW $\text{m}^{-1} K^{-1}$ ) 
        &$c_v$ ($kJ kg^{-1} K^{-1}$) & $\gamma = \frac{c_p}{c_v}$ & Eq.~\eqref{eq:inequality-ideal-gas} satisfied? \T\B \\ \hline
        Water at 274 K, 1 bar & 1791.0  & 561.09 &4.2170  & 1.00057 & \checkmark
       \T\B \\ \cline{1-6}
       Para-Hydrogen at 20 K , 1 bar & 13.63 & 103.1 &
       5.637 & 1.67 &\checkmark \T\B \\ \cline{1-6}
       Propane at 85.53 K, 1 bar & 10790. & 207.9 &1.355 &1.414 & \xmark \T\B \\ \cline{1-6}
       Ethane at 90.38 K, 1 bar & 1281. &255.6 & 1.605 & 1.44& \xmark \T\B \\ \hline
    \end{tabular}
    \label{tab:liquids}
\end{table*}

We now check if the prediction from causality and reduction to the Navier-Stokes equations
[Eq.~\eqref{eq:inequality-thermal-conductivity}] 
is satisfied for experimental and theoretical calculations available for some liquids and gases.
Let us begin by considering a fluid that can be modeled as an ideal gas.
For an ideal gas with adiabatic index $\gamma$, we have
\begin{subequations}\label{eq:ideal_gas_eos}
\begin{align}
    p = &
    (\gamma -1) \rho \epsilon
    \,,\\
    \epsilon 
    = & 
    c_v T 
    \,,\\
    c_v = &
    \frac{k_B}{m \left(\gamma-1 \right)}
    \,.
\end{align}
\end{subequations}
Using this in Eq.~\eqref{eq:inequality-thermal-conductivity}, we find
\begin{align}\label{eq:inequality-ideal-gas}
    \frac{\kappa^{(0)}}{\eta^{(0)}} > \frac{k_B}{m}
    \implies  f\equiv \frac{\kappa^{(0)}}{\eta^{(0)} c_v} > \gamma -1\,,
\end{align}
where, $c_v$ is the specific heat at constant volume.

Let us next check when this inequality is satisfied with theoretical predictions available from Chapman-Enskog theory.
Chapman-Enskog theory predicts a simple relationship between $\kappa^{(0)}$ and $\eta^{(0)}$
\begin{align}
    \kappa^{(0)} = f \eta^{(0)} c_v
\end{align}
where $f$ is a purely numerical factor that is theoretically predicted to lie between $2.5$ to $2.522$ 
for many microscopic models\footnote{The details of this calculation are available in 
Chapters 7-11 of~\cite{Chapman-Cowling-Book} and a summary is provided in Chapters 12-13.}. 
For mono-atomic gases, $\gamma = 5/3$, and Eq.~\eqref{eq:inequality-ideal-gas} predicts that
\begin{align}
    f > 2/3 \sim 0.67\,.
\end{align}
This inequality is satisfied for both the theoretical value $f \sim 2.5$ 
and for several experimentally observed values; see Table 22 of~\cite{Chapman-Cowling-Book}.
For diatomic gases, $\gamma = 7/5$ and Eq.~\eqref{eq:inequality-ideal-gas} 
predicts that $f> 0.4$ which is satisfied by the experimentally observed value of 
$f = 1.9$~\cite{Chapman-Cowling-Book}.
Therefore, calculations available from kinetic theory satisfy this inequality.

We also compare the inequality with the properties of some common liquids in Table~\ref{tab:liquids}. 
We find that Eq.~\eqref{eq:inequality-thermal-conductivity} is violated for many common liquids. 
This violation could be explained in several ways.
First, we have assumed an ideal gas equation of state.
Second, we have assumed that the fluids in question could be described as a fluid at both
high (relativistic) and low (non-relativistic) energy scales.
Finally, we have assumed the fluids could be modeled as a BDNK (first-order) fluid.
A more detailed analysis will be required to determine when Eq.~\eqref{eq:upper-bound-eta-by-s} 
could be applied to non-relativistic fluids.
\subsection{Discussion}\label{subsec:Discussion}
We derived Eq.~\eqref{eq:inequality-thermal-conductivity} 
by asserting that the non-relativistic limit reduces exactly 
to the Navier-Stokes equations.
We now show that one could derive the same inequality  
if one assumes that 
\begin{align}
    \tau_{\epsilon v}^{(-2)} \tau_{\rho , \perp}^{(0)} \rho \ll \tau_{\epsilon,\perp} \left(\zeta^{(0)} + \frac{\eta^{(0)}}{3} \right)\,.
\end{align}
This assumption implies that the corrections to the Navier-Stokes 
equation arising from the right-hand side of Eq.~\eqref{eq:NR-limit-eqs-general} are small even if the gradient corrections are of order unity. 
We expand $a_3 >0$ and obtain
\begin{align}
   a_3 
   &=  
   -
   \tau_{\epsilon,\perp}^{(0)} \left( \zeta^{(0)} + \frac{\eta^{(0)}}{3}\right) 
   + 
   \tau_{\epsilon v}^{(-2)} \tau_{\rho,\perp}^{(0)} \rho 
   + 
   \mathcal{O}(c^{-2}) 
   \nonumber 
   \,, \\
   &\approx 
   -
   \tau_{\epsilon,\perp}^{(0)} \left( \zeta^{(0)} + \frac{\eta^{(0)}}{3}\right) 
   + 
   \mathcal{O}(c^{-2}) 
   >
   0 
   .
\end{align}
With this, we can obtain the inequality on $\eta^{(0)}/s$ 
by following the same steps followed in Eq.~\eqref{eq:a_3-geq-NS-simplification}. One can of course also ensure that the BDNK corrections are small by focusing only on solutions that are ``smooth
enough,'' such that the higher-order gradient terms are intrinsically small. Doing so, however, would severely limit the applicability of BDNK fluids.  

How should we interpret the non-relativistic limit of the 
relativistic, viscous fluid equations of motion?
The Navier-Stokes equations are highly successful in modelling fluid flow in a number of situations, such as 
oceanography, turbulence, weather prediction and even in astrophysical environments, such as accretion disks~\cite{batchelor_2000}. This success occurs when describing fluids with both mild and steep gradients, provided the non-relativistic expansion assumptions hold. 
Therefore, one must ensure that the first-order relativistic fluid model of interest 
reduces to the Navier-Stokes equations in the non-relativistic limit. 
The question then is whether this reduction should be \textit{exact} or \textit{approximate} (in an expansion in gradients of the fluid solution). 
Experiment could decide this question, but given the success of the Navier-Stokes system, 
we expect the size of the corrections arising from the right-hand side of Eq.~\eqref{eq:NR-limit-eqs-general} to be small.
If this is true, then the new BDNK coefficients should be small compared
to the other dynamical times for the non-relativistic equations of motion,
which implies that the inequality of Eq.~\eqref{eq:inequality-thermal-conductivity}
should still approximately hold.

\section{Non-relativistic limit of an extended variable model}\label{sec:MIS-model}
In this section, we compare our results with an extended variable model that has been shown to be strongly hyperbolic and causal~\cite{Bemfica:2019cop}.
The main result of this section is that unlike first-order relativistic fluid models, 
the non-relativistic equations of motion for extended variable models can be
hyperbolic or elliptic, depending on the scaling of the coefficients with $1/c$. 

The stress-energy tensor and the conservation current for the model we consider are given by
\begin{align}
    T_{\mu\nu} &= 
    e\frac{u_{\mu} u_{\nu}}{c^2}
    +
    \left(p + \Pi \right) \Delta_{\mu \nu}\,, \\
    J^{\mu} &= \rho u^{\mu}\,,
\end{align}
where $\Pi$ is a new dynamical field,
which obeys the relaxation equation
\begin{align}
    \label{eq:mis-transport-eqn-Pi}
    \tau_{\Pi} u^{\alpha} \nabla_{\alpha} \Pi 
    + 
    \lambda \Pi^2 
    + 
    \Pi 
    + 
    \zeta \theta 
    &=0
    \,,
\end{align}
$\tau_{\Pi}$ is a relaxation time and $\lambda$ is another transport coefficient.
Unlike the BDNK fluid model, 
this model only incorporates the effects of bulk viscosity $\zeta$.
Besides the new equation of motion \eqref{eq:mis-transport-eqn-Pi}, the other equations
of motion arise from the same equations as those of first-order models,
$\nabla_{\mu}T^{\mu\nu}=0$ and $\nabla_{\mu}J^{\mu}=0$.
This model was shown to be hyperbolic and causal if~\cite{Bemfica:2019cop}
\begin{align}\label{eq:causality-constraint-MIS}
    \left[
        \frac{\zeta}{\tau_{\Pi}} 
        + 
        \rho \partial_{\rho} p 
    \right] 
    \frac{1}{e + p + \Pi}
    \leq 
    1 
    - 
    \frac{\partial_{\epsilon} p }{\rho}
    \,.
\end{align}

Studying the non-relativistic limit of this model is relatively straightforward.
We first expand the transport coefficients as
\begin{align}
    \tau_{\Pi} 
    &= 
    \tau_{\Pi}^{(0)} 
    + 
    \frac{\tau_{\Pi}^{(2)} }{c^2} 
    + 
    \mathcal{O}(c^{-4}) 
    \,, \\
    \lambda 
    &= 
    \lambda^{(0)} 
    + 
    \frac{\lambda^{(2)} }{c^2} 
    + 
    \mathcal{O}(c^{-4}) 
    \,.
\end{align} 
We then expand the non-relativistic equations as
\begin{subequations}
\begin{align}
    D^N \rho 
    + 
    \rho \theta^N 
    &= 
    \mathcal{O}\left(c^{-2}\right) 
    \,, \\
    \label{eq:Deps-MIS-Newt-v1}
    \rho D^N \epsilon 
    + 
    \left(p + \Pi\right) \theta^N  
    &=  
    \mathcal{O}\left(c^{-2}\right)
    \,,\\
    \label{eq:Dv-MIS-Newt-v1}
    \rho D^N v^j 
    +
    \partial^j \left(p + \Pi\right)
    &=  
    \mathcal{O}\left(c^{-2}\right) 
    \,, \\
    \label{eq:tau-Pi-eqs-Newt-general}
    \tau_{\Pi}^{(0)} D^N \Pi 
    + 
    \lambda^{(0)} \Pi^2 
    +
    \Pi 
    + 
    \zeta^{(0)} \theta^N 
    &= 
    \mathcal{O}\left(c^{-2}\right) 
    \,.
\end{align}
\end{subequations}
One can follow the methods in~\cite{Bemfica:2019cop} to show that the above equations are strongly-hyperbolic and causal if certain constraints are satisfied~\cite{jorge_upcoming}.
In order to force the non-relativistic equations of motion to reduce to the Navier-Stokes equations,
we set $\tau_{\Pi}^{(0)} = 0$ and $\lambda^{(0)} = 0$.
When this holds, Eq.~\eqref{eq:tau-Pi-eqs-Newt-general} reduces to
\begin{align}
    \label{eq:nr-limit-Pi}
    \Pi 
    + 
    \zeta^{(0)} \theta^{N} 
    &= 
    \mathcal{O}\left(c^{-2}\right)
    \,.
\end{align} 
When we then substitute Eq.~\eqref{eq:nr-limit-Pi}
into Eqs.~\eqref{eq:Deps-MIS-Newt-v1} and \eqref{eq:Dv-MIS-Newt-v1},
we obtain the Navier-Stokes equations with bulk viscosity
\begin{subequations}
\label{eq:nr-ns-just-zeta-israel-stewart}
\begin{align}
    D^N \rho 
    + 
    \rho \theta^N 
    &= 
    \mathcal{O}\left(c^{-2}\right) 
    \,, \\
    \rho D^N \epsilon 
    + 
    \left(p -\zeta^{(0)} \theta^N\right) \theta^N  
    &=  
    \mathcal{O}\left(c^{-2}\right)
    \,,\\
    \rho D^N v^j 
    + 
    \partial^j \left(p - \zeta^{(0)} \theta^N\right)
    &=  
    \mathcal{O}\left(c^{-2}\right) 
    .
\end{align}
\end{subequations}
As the Navier-Stokes equations are parabolic \cite{kreiss_oliger_ns_eqns_book}, 
we conclude that the above system \eqref{eq:nr-ns-just-zeta-israel-stewart} is not hyperbolic.

Finally, again setting $\tau_{\Pi}^{(0)}=0$ and $\lambda^{(0)}=0$, 
we expand the causality constraint of Eq.~\eqref{eq:causality-constraint-MIS}, and obtain
\begin{align}
    \label{eq:nr-causality-constraint-MIS}
    \frac{ \zeta^{(0)}}{\rho \tau_{\Pi}^{(2)}} 
    \leq 
    1 
    - 
    \frac{\partial_{\epsilon} p }{\rho} 
    + 
    \mathcal{O}\left(c^{-2}\right)
    .
\end{align}
We see that Eq.~\eqref{eq:nr-causality-constraint-MIS}
gives us an upper bound on the value of bulk viscosity 
in the non-relativistic regime.
If we had instead kept $\tau_{\Pi}^{(0)}\neq0$, 
the non-relativistic limit of Eq.~\eqref{eq:causality-constraint-MIS} would have given
us $0\leq 1 - \partial_{\epsilon}p/\rho$, 
which would restrict the kinds of equation of state that could be used with this model.

\section{Conclusions}\label{sec:conclusions}
The BDNK model is the first relativistic fluid model that has been shown 
to be causal, to have a locally well-posed initial value problem, 
and to have modally stable equilibrium solutions in flat space.
The hyperbolicity and modal stability of the BDNK fluid depends on the parametric
form of the transport coefficients (i.e.,~the choice of hydrodynamic frame).
Finding suitable hydrodynamics BDNK frames has proven to be a challenge to 
understand the physical predictions of BDNK fluids.
To better understand the nature of different BDNK frames, 
in this article we studied the non-relativistic limit of the BDNK model,
and showed that the Newtonian limit of the causality constraints
can constrain the value of physical transport coefficients, such as the shear viscosity.
Moreover, we showed that the only known BDNK frame for a non-conformal fluid
\cite{Pandya:2022sff} 
does not limit to a viscous fluid in the non-relativistic limit, nor
does it satisfy preservation of the structure of the post-Newtonian metric 
(see Sec.~\ref{sec:NR-limit-BDNK}). 

We showed that the non-relativistic limit of any first-order 
fluid model always reduces to a non-hyperbolic theory.
Next, we analyzed the conditions under which the non-relativistic equations 
[Eq.~\eqref{eq:NR-limit-eqs-general}] 
reduce \emph{exactly} to the Navier-Stokes equations.
This analysis, combined with the causality/hyperbolicity constraints,
shows that hyperbolic first-order fluid models satisfy the following:
\begin{enumerate}
    \item The non-relativistic Fourier-law is always corrected 
    by higher-order gradients [Eq.~\eqref{eq:heat-flux-BDNK-NR}].
    \item With the additional assumption that the non-relativistic limit of
    the momentum equation reduces \emph{exactly} to the Navier-Stokes equation,
    or alternatively that the non-hydrodynamic modes decay much faster than
    the hydrodynamic modes,
    the ratio of the thermal conductivity to the shear 
    viscosity of the fluid satisfies an inequality
    [Eq.~\eqref{eq:inequality-thermal-conductivity}].
    This inequality can alternatively be written in terms of the
    ratio of the shear viscosity to the entropy density.
\end{enumerate}
As a point of comparison, we then expanded a simple 
hyperbolic extended-variable model~\cite{Bemfica:2019cop} in the non-relativistic limit.
We showed that this limit forces the extended variable model to reduce to either a hyperbolic theory or a non-hyperbolic theory, 
depending on the scaling in $1/c$ of the transport coefficients.

There are several directions for future work.
First, new BDNK hydrodynamical frames for non-conformal 
fluids--which are of the most direct (astro)physical significance--should be found. 
In particular, both relativistic kinetic theory \cite{Israel-1963,Stewart-1971,cercignani2002relativistic} 
and the preservation of the structure of the post-Newtonian metric condition require that the BDNK transport coefficients
scale as \emph{even} powers of the speed of light $c$.
Finding any non-conformal frame that respects this property, and that satisfies
the causality and stability constraints, remains an unsolved problem for
BDNK fluids.

Another direction for future work is to find the non-relativistic limit of the
equations of motion for other fluid models, 
such as the ones derived from the DNMR formalism~\cite{Denicol_2012}, 
or the rBRSSS model~\cite{Baier:2007ix}.
These extended-variables models have been used to describe heavy-ion collisions, 
but mathematical results, such as 
local well-posedness and causality, have only been recently derived in some 
restrictive cases~\cite{Bemfica:2019cop,Bemfica:2020xym}.
These models are essentially generalizations of MIS models, and have a 
large number of free parameters that have
to be fit to observational/experimental data. 
Such a large number of parameters introduces the potential problem of over-fitting observed data\footnote{The Newtonian 
MIS models are known to have many parameters, 
but they do not fit observations of fluid flow better than Navier-Stokes predictions, 
unless one includes hundreds of new parameters~\cite{Muller-book}.}.
Therefore, studying the non-relativistic limit and the constraints imposed by 
causality could be very beneficial to understand the different transport coefficients 
for those theories.

In this work, we did not consider the most general first-order 
fluid stress-energy tensor one could write down,
as we did not add gradient corrections to the fluid rest-mass current (see Eqs.~\eqref{eq:J-decomp}, \eqref{eq:N-def-BDNK} and \eqref{eq:diffusion-current-BDNK-def}).
One possible avenue for future work could be to determine the non-relativistic limit of this general model and study the implications 
of the causality constraints for non-relativistic motion.

Finally, another interesting avenue is to check the robustness of our upper bound on
$\eta^{(0)}/s$ [Eq.~\eqref{eq:upper-bound-eta-by-s}].
As we verified in Sec.~\ref{subsec:NR-limit-NS-condition},
the bound is satisfied by calculations available from Chapman-Enskog theory.
It would be interesting to derive this upper bound from microscopic arguments.
Moreover, as we showed in Eq.~\eqref{eq:lower-bound-kappa-KSS} 
one can use this upper bound, along with the KSS conjecture, to derive a lower bound on 
$\rho \kappa^{(0)} s^{-2}$.
Perhaps, this bound can also be derived from the AdS-CFT conjecture.
\begin{acknowledgements}
We acknowledge support from the Simons Foundation through Award number 896696 and the NSF through award PHY-2207650. 
We thank Jorge Noronha, Marcelo Disconzi and Charles Gammie for many insightful discussions that improved our presentation and allowed us to generalize our results.
We thank Luis Lehner, Elias Most, Alex Pandya, Jacquelyn Noronha-Hostler and Helvi Witek for discussions regarding this work.

\end{acknowledgements}
\appendix
\section{Definitions and units}\label{appendix:units}
For convenience, we collect the definitions and SI units of the 
constants and fields we make use of in this article in Table~\ref{tab:units}.

We take as our fundamental fluid variables $u^{\mu},\rho,\epsilon$. 
From the equation of state, one can determine all of the other fluid variables
from $\rho,\epsilon$,
for example $\mu\left(\rho,\epsilon\right)$, $\zeta\left(\rho,\epsilon\right)$.
\begin{table*}[thp]
\caption{List of definitions and units}
\centering
    \begin{tabular}{|c|c|c|}
        \hline
        Quantity & 
        Definition & 
        Units (SI)
        \\
        \hline
        Speed of light & 
        $c$ & 
        $\mathrm{m}\times\mathrm{s}^{-1}$ 
        \\
        \hline
        Fluid four-velocity & 
        $u^{\mu}$ & 
        $\mathrm{m}\times\mathrm{s}^{-1}$ 
        \\
        \hline
        Particle density & 
        $n$ & 
        $\mathrm{m}^{-3}$ 
        \\
        \hline
        Particle mass & 
        $m_b$ & 
        $\mathrm{kg}$ 
        \\
        \hline
        Rest-mass density & 
        $\rho = m_bn$ & 
        $\mathrm{kg}\times\mathrm{m}^{-3}$ 
        \\
        \hline
        Total energy density & 
        $e = \rho \left(c^2 + \epsilon\right)$ & 
        $\mathrm{kg}\times\mathrm{m}^{-1}\times\mathrm{s}^{-2}$ 
        \\
        \hline
        Specific internal energy & 
        $\epsilon$ & 
        $\mathrm{m}^{2}\times\mathrm{s}^{-2}$ 
        \\
        \hline
        Specific enthaply & 
        $h = c^2 + \epsilon + p/\rho$ & 
        $\mathrm{m}^{2}\times\mathrm{s}^{-2}$ 
        \\ 
        \hline
        Pressure & 
        $p$ & 
        $\mathrm{kg}\times\mathrm{m}^{-1}\times\mathrm{s}^{-2}$ 
        \\
        \hline
        Chemical potential & 
        $\mu$ & 
        $\mathrm{kg}\times\mathrm{m}^2\times\mathrm{s}^{-2}$ 
        \\
        \hline
        Temperature & 
        $T$ & 
        $\mathrm{K}$ 
        \\
        \hline
        Fugacity & 
        $\varphi = \mu/T$ & 
        $\mathrm{kg}\times\mathrm{m}^2\times\mathrm{s}^{-2}\times\mathrm{K}^{-1}$ 
        \\
        \hline
        Bulk viscosity & 
        $\zeta$ & 
        $\mathrm{kg}\times\mathrm{m}^{-1}\times\mathrm{s}^{-1}$ 
        \\
        \hline
        Shear viscosity & 
        $\eta$ & 
        $\mathrm{kg}\times\mathrm{m}^{-1}\times\mathrm{s}^{-1}$ 
        \\
        \hline
        Thermal conductivity & 
        $\kappa$ & 
        $\mathrm{kg}\times\mathrm{m}\times\mathrm{s}^{-3}\times\mathrm{K}^{-1}$ 
        \\
        \hline
    \end{tabular}
\label{tab:units}
\end{table*}
\section{Principal symbol for a general parameterization}\label{appendix:principal-symbol-general}
Here calculate the characteristics of the system for the relativistic fluid equations, 
[Eqs.~\eqref{eq:j-mu-conservation} and \eqref{eq:t-mu-nu-conservation}], 
given the parametrization \eqref{eq:parameterization-general}.
To compute the characteristics, we first write down the
principal part of the equations of motion 
(for a review of the relevant terminology, see e.g. \cite{kreiss_oliger_ns_eqns_book}).
After some algebra, we have
\begin{subequations}\label{eq:principal-part-general-equations}
\begin{align}
    \label{eq:principal-part-conservation-current}
    &
    u^{\alpha}u^{\beta} \partial_{\alpha}\partial_{\beta} \rho
    + 
    \rho \delta^{(\alpha}_{\nu} u^{\beta)} \partial_{\alpha}\partial_{\beta} u^{\nu} 
    +
    \text{l.o.t} =0\,,\\
    \label{eq:principal-part-energy}
    &\left(\frac{\tau_{\epsilon,\parallel}}{c^2} u^{\alpha} u^{\beta} 
    +
    \frac{\tau_{\epsilon,\perp}}{c^2} \Delta^{\alpha \beta} 
    \right)\partial_{\alpha}\partial_{\beta}
     \epsilon
    +
    \frac{\tau_{\rho,\perp}}{c^2} \Delta^{\alpha \beta} \partial_{\alpha}\partial_{\beta} \delta \rho
    \nonumber \\
    &+
    \frac{\tau_{u}}{c^2}  u^{(\alpha} \delta^{\beta)}_{\nu} \partial_{\alpha}\partial_{\beta}
    u^{\nu} +\text{l.o.t}
    =0 \,,\\
    \label{eq:principal-part-momentum}
    &\frac{\tau_{\epsilon v}}{c^2} \Delta^{\mu(\alpha} u^{\beta)}\partial_{\alpha}\partial_{\beta} 
     \epsilon
    +
    \frac{\tau_{\rho, \perp}}{c^2} \Delta^{\mu(\alpha} u^{\beta)}\partial_{\alpha}\partial_{\beta} 
     \rho
    \nonumber \\
    &+
    \hat{C}^{\mu \alpha \beta}_{\nu} \partial_{\alpha}\partial_{\beta}
     u^{\nu}
    +\text{l.o.t}
    =0\,,
\end{align}
\end{subequations}
where,
\begin{align}
    \hat{C}^{\mu \alpha \beta}_{\nu}
    &=
    \left(V_1 + \eta + \frac{\rho \tau_{\rho,\perp}}{c^2}\right)
    \Delta^{\mu(\alpha} \delta^{\beta)}_{\nu} \nonumber\\
    &+ 
    \left( 
    \frac{\tau_{Q,1}}{c^2} u^{\alpha}u^{\beta}
    -
    \eta \Delta^{\alpha \beta}
    \right)
    \delta^{\mu}_{\nu}
    \,.
\end{align}
and $\text{l.o.t.}$ stands for lower order terms (in derivatives) in the equations of motion.
To derive \eqref{eq:principal-part-conservation-current}, we computed the principal part of
$u^{\alpha}\nabla_{\alpha}\nabla_{\beta}J^{\beta}=0$.
To derive \eqref{eq:principal-part-energy}, we computed the principal part of the energy
equations, and we made use of \eqref{eq:principal-part-conservation-current}
to replace $u^{\alpha}u^{\beta}\partial_{\alpha}\partial_{\beta}\rho$ with 
$-\rho \delta^{(\alpha}_{\nu} u^{\beta)} \partial_{\alpha}\partial_{\beta} u^{\nu}$.
Equation~\eqref{eq:principal-part-momentum} constitutes principal part of the momentum equation.
To derive this equattion, 
we again made use of \eqref{eq:principal-part-conservation-current} 
to replace $u^{\alpha}u^{\beta}\partial_{\alpha}\partial_{\beta}\rho$ with 
$-\rho \delta^{(\alpha}_{\nu} u^{\beta)} \partial_{\alpha}\partial_{\beta} u^{\nu}$.

One can then follow the same approach as in~\cite{Bemfica:2020zjp} 
to calculate the characteristics of the above system.
Our analysis closely follows Appendix A of~\cite{Bemfica:2020zjp}.
Given a co-vector $\xi_{\alpha}$ define
\begin{align}
    \xi_{\alpha} &= -\frac{b}{c^2} u_{\alpha} + v_{\alpha} \,,\\
    b&\equiv \xi_{\alpha} u^{\alpha} \,,\\
    v^{\alpha} &\equiv \Delta^{\alpha \beta} u_{\beta}\,.
\end{align}
The principal symbol is obtained by replacing $\partial_{\alpha} \to \xi_{\alpha}$ in the principal part of the differential equations given in Eq.~\eqref{eq:principal-part-general-equations}.
The principal symbol is then
\begin{align}
    &\mathcal{M} 
    = 
    \nonumber \\ 
    \label{eq:principal-symbol-general}
    &\begin{pmatrix}
        0 & b^2 & \rho b \xi_{\nu} \\
        \frac{\tau_{\epsilon,\parallel}}{c^2} b^2 + \frac{\tau_{\epsilon,\perp}}{c^2} v^{\alpha} v_{\alpha}
        & \frac{\tau_{\rho,\perp}}{c^2}  v^{\alpha} v_{\alpha}
        & \frac{\tau_u}{c^2} b \xi_{\nu} \\
        \frac{\tau_{\epsilon v}}{c^2} b v^{\mu} & \frac{\tau_{\rho,\perp}}{c^2} b v^{\mu} & 
        \mathcal{D}^{\mu}_{\nu}
    \end{pmatrix}
\end{align}
where
\begin{align}
    \mathcal{D}^{\mu}_{\nu} 
    \equiv&
    \hat{C}^{\mu \alpha \beta}_{\nu} \xi_{\alpha}\xi_{\beta} 
    \,\nonumber\\
    =& 
    v^{\mu}\xi_{\nu} 
    \left( 
        V_1 + \eta + \frac{\rho \tau_{\rho,\perp}}{c^2}
    \right)
    \nonumber\\
    &
    +
    \left( 
        \frac{\tau_{Q,1} b^2}{c^2}
        -
        \eta v_{\alpha}v^{\alpha}
    \right)
    \delta^{\mu}_{\nu}
\end{align}
The characteristic speeds $\mathfrak{c}$ are given by solutions to 
$\text{det} \mathcal{M} = 0$ with
\begin{align}
    \mathfrak{c}^2 = \frac{b^2}{v^2}\,.
\end{align}
The determinant of the principal symbol [Eq.~\eqref{eq:principal-symbol-general}] 
can be found using Eq.~\eqref{eq:Z-det}, which gives us
\begin{align}
    &
    \text{det} \mathcal{M} 
    =
    \nonumber\\
    &
    - 
    \frac{b^2}{c^2} \left[\frac{\tau_{Q,1} b^2}{c^2} - \eta v^2 \right]^3 \left[ 
    a_1 \frac{b^4}{c^4} 
    -
    a_2 \frac{b^2}{c^2} v^2
    +
    a_3 v^4
    \right]\,.
\end{align}
where, $a_1,a_2$ and $a_3$ are defined in Eq.~\eqref{eq:ai-defs} of the main text.
We see that $\text{det}\mathcal{M}=0$ when
\begin{subequations}
\begin{align}
   \mathfrak{c}^2
   =&
   0
   ,\\
   \frac{\mathfrak{c}^2}{c^2}
   =&
   \frac{\eta}{\tau_{Q,1}}
   ,\\
   \frac{\mathfrak{c}^2}{c^2}
   =&
   \frac{a_2\pm\sqrt{a_2^2-4a_1a_3}}{2a_1}
   .
\end{align}
\end{subequations}
For hyperbolicity we require $0\leq\mathfrak{c}^2$
(that is, $\mathfrak{c}$ must be real), 
and for causality we require $\mathfrak{c}^2/c^2\leq 1$.
These conditions give the causality constraints Eq.~\eqref{eq:causality-general-model} quoted in the main text.
Demonstrating that the equations of motion have a well-posed initial value problem requires
showing that the principal symbol has a complete set of eigenvectors in addition to having all
real eigenvalues (the equations of motion are \emph{strongly hyperbolic}). 
We note that as \eqref{eq:principal-symbol-general} takes the same form as the principal symbol
does in \cite{Bemfica:2020zjp}, we can use the same results invoked in Appendix C of that
reference to conclude the equations of motion are strongly hyperbolic, and hence
have a well-posed initial value problem, provided the causality constraints are satisfied.
\section{Details of the non-relativistic expansion of the BDNK model}\label{appendix:details-nr-expansion}
\subsection{Non-relativistic limit}\label{appendix:PN-identities-Mink}
In this appendix, we provide the details of the derivation carried out in Sec.~\ref{sec:NR-limit-BDNK}.
We first list some useful identities. 
From the definitions of the total energy density $e$ and the chemical potential $\mu$ we see that
\begin{align}
    \label{eq:e-def-PN}
    \frac{e}{c^2} &= \rho + \frac{\rho \epsilon}{c^2} \,,\\
    \label{eq:mu-def-PN}
    \frac{\mu}{c^2} &= m_b + \frac{\mu_N}{c^2}\,.
\end{align}
We next expand $W$,$D$, $\theta$, and the components of $\Delta^{\alpha \beta}\nabla_{\beta}$
\begin{align}
    W &= 1 +\frac{v^2}{2c^2} + \frac{3 v^4}{8 c^4} + \mathcal{O}(c^{-6})\,, \\
    \label{eq:D-expansion-NR}
    D
    &= 
    \left(
        1 
        + 
        \frac{v^2}{2c^2} 
    \right)D^N + \mathcal{O}\left(c^{-4}\right)
    ,\\
    \label{eq:theta-expansion-NR}
    \theta 
    &= 
    \theta^N 
    + 
    \frac{1}{2c^2}\left(D^N \left(v^2\right) + \theta^N v^2 \right)\,, \nonumber \\
    &+ \mathcal{O}\left(c^{-4}\right)\,
    \\ 
    \label{eq:Delta-0-beta-nabla-beta-expansion-NR}
    \Delta^{0\beta}\nabla_{\beta} 
    &=
    \frac{1}{c} v^i \partial_i + \frac{v^2}{ c^3} D^N + \mathcal{O}\left( c^{-5}\right)
    ,\\
    \label{eq:Delta-i-beta-nabla-beta-expansion-NR}
    \Delta^{i\beta}\nabla_{\beta}
    &=
    \partial_i
    +
    \frac{v^2}{c^2}
    v^i D^N
    +
    \mathcal{O}\left(c^{-4}\right)\,.
\end{align}
We have used the Newtonian convective derivative $D^N\equiv\partial_t+v^i\partial_i$.
The expansion of the components of the projection operator $\Delta^{\alpha\beta}$ are
\begin{align}
    \label{eq:Delta-00-expansion-NR}
    \Delta^{00}
    &=
    \frac{v^2}{c^2} + \mathcal{O}\left(c^{-4}\right)\,,
    \\
    \label{eq:Delta-0-i-expansion-NR}
    \Delta^{0i}
    &=
    \frac{v^i}{c} + \frac{v^i v^2}{c^3} +\mathcal{O}\left(c^{-5}\right)\,,
    \\
    \label{eq:Delta-i-j-expansion-NR}
    \Delta^{ij}
    &=
    \delta^{ij}
    +
    \frac{v^iv^j}{c^2}
    + \mathcal{O}\left(c^{-4}\right)
    .
\end{align}
We use these to expand the decomposition of the gradient of
the fluid velocity.
The components of the acceleration vector are given by (remember that the
Christoffel symbols are zero for the metric $\eta_{\alpha \beta}$)
\begin{align}
    \label{eq:a-0-PN-mink}
    a^0 
    &= 
    D^N\left(\frac{v^2}{2c} \right) 
    + 
    \frac{1}{c^3} \left( 
        \frac{3}{8} D^N\left(v^4\right) 
        + 
        \frac{v^2}{4} D^N\left(v^2\right)
    \right) \nonumber \\
    &+ 
    \mathcal{O}\left(c^{-5}\right)\,,
    \\
 \label{eq:a-j-PN-mink}
    a^j 
    &= 
    D^N v^j 
    + 
    \frac{1}{c^2} \left( 
        \frac{v^2 D^N v^j +  D^N (v^2 v^j)}{2}  
    \right) 
    \nonumber \\
    &+ 
    \mathcal{O}\left(c^{-4}\right)\,.
\end{align}
The components of the shear tensor are given by
\begin{align}
 \sigma^{00} 
    &= 
    \mathcal{O}\left(c^{-2}\right)
    \,, 
    \\
    \sigma^{0j} 
    &= 
    \frac{v_k}{c} \left(\sigma^N\right)^{jk}
    +
    \mathcal{O}\left(c^{-3}\right)
    \,,\\
    \sigma ^{ij} 
    &= 
    \left(\sigma^N\right)^{ij}
    +
    \mathcal{O}\left(c^{-2}\right)\,,
\end{align}
where $\sigma^N_{ij} \equiv \partial_{(i}v_{j)} - \frac{1}{3}\delta_{ij}\theta^N$.
The contraction of the shear tensor gives us
\begin{align}
    \sigma_{\mu \nu}\sigma^{\mu \nu} 
    &= 
    \sigma^N_{ij}\left(\sigma^N\right)^{ij}
    + 
    \mathcal{O}\left(c^{-2}\right)
    \,.
\end{align}

The non-relativistic expansion of $\mathcal{E}$ (see Eq.~\eqref{eq:E-def-general-param}) is
\begin{align}
\label{eq:PN_expansion_E_mink}
        \mathcal{E}
        &=
        \rho +
        \frac{1}{c^2}
        \left(
            \rho\epsilon 
            +
            \tau_{\epsilon,1}^{(0)} D^N \epsilon
            +
            \tau_{\epsilon,2}^{(0)} \theta^N
            +
            \tau_{\epsilon,3}^{(0)} D^N \left(\frac{m_b}{T} \right)
        \right)
        \nonumber\\
        &+
        \mathcal{O}\left(c^{-4}\right)\,,
\end{align}
where we have used the non-relativistic expansion of the transport coefficients in 
Eq.~\eqref{eq:nr-expansion-transport-coeffs}, along with
Eqs.~\eqref{eq:e-def-PN}-\eqref{eq:mu-def-PN}.
The non-relativistic expansion of $\mathcal{P}$ (see Eq.~\eqref{eq:P-def-general-param}) is
\begin{align}
\label{eq:PN_expansion_P_mink}
        \mathcal{P}
        &=
        p 
        + \tau_{p,1}^{(0)} D^N \epsilon
        + (\tau_{p,2}^{(0)}-\zeta^{(0)}) \theta^N
        + \tau_{p,3}^{(0)} D^N \left( \frac{m_b}{T}\right) \nonumber\\
        &+
        \mathcal{O}\left(c^{-2}\right)\,,
\end{align}
where, we have used the non-relativistic expansion of the transport coefficients in 
Eq.~\eqref{eq:nr-expansion-transport-coeffs} and Eq.~\eqref{eq:mu-def-PN}.

We next consider the non-relativistic expansion of $\mathcal{Q}^{\mu}$,
which is defined in Eq.~\eqref{eq:Q-def-general-param}.
The $t$ component of the heat vector is given by
\begin{align}
     \mathcal{Q}^0 
    &= \frac{1}{c} \left[ 
    \frac{\tau_{Q,1}^{(-2)}}{\rho} v_j E^j
    \right]
     + \mathcal{O}(c^{-3})\,, \nonumber\\
    \label{eq:Q-0-PN-Mink-2-def}
    &
    \equiv\frac{\mathcal{Q}^{0}_{1}}{c} + \mathcal{O}(c^{-3})\,.
\end{align}
We have used the PN expansion of $a^0$ [Eq.~\eqref{eq:a-0-PN-mink}], $\Delta^{0 \alpha} \nabla_{\alpha}$ [Eq.~\eqref{eq:Delta-0-beta-nabla-beta-expansion-NR}], $\mu$ [Eq.~\eqref{eq:mu-def-PN}] and $e$ [Eq.~\eqref{eq:e-def-PN}] to obtain the first line.

The spatial component of the heat vector can be obtained by a similar calculation using the PN expansion of $a^j$ [Eq.~\eqref{eq:a-j-PN-mink}], $\Delta^{j \alpha} \nabla_{\alpha}$ [Eq.~\eqref{eq:Delta-i-beta-nabla-beta-expansion-NR}], $\mu$ [Eq.~\eqref{eq:mu-def-PN}] and $e$ [Eq.~\eqref{eq:e-def-PN}].
The final answer is given by
\begin{align}
    \mathcal{Q}^j 
    &=\frac{\tau_{Q,1}^{(-2)}}{\rho} E^j 
    \!+\!
    \frac{1}{c^2} \left[
        \frac{ \tau_{Q,1}^{(-2)} \left( v^2 D^N v^j + D^N \left( v^2 v^j\right) \right)}{2\rho} \right. \nonumber \\
        &+ \left.
        \frac{\tau_{Q,1}^{(-2)}}{\rho}\left(\rho \epsilon + p\right) D^N v^j 
        +
       \frac{\tau_{Q,1}^{(-2)}}{\rho} v^2 v^j D^N p  
        \right.
        \nonumber \\
        &\left.
        - 
        \kappa^{(0)} \partial_j T 
        + \left[\frac{\tau_{Q,1}^{(0)}}{\rho} - \frac{\epsilon + p }{\rho^2} \tau_{Q,1}^{(-2)} \right] E^j
    \right] + \mathcal{O}(c^{-4})\,,\nonumber \\
    \label{eq:Q-j-PN-Mink-2-def}
    &\equiv 
    \mathcal{Q}^{j}_{0} + \frac{\mathcal{Q}^{j}_{2}}{c^2} + \mathcal{O}(c^{-4})\,.
\end{align}
We also need the non-relativistic expansions of the following quantities
\begin{align}
\label{eq:theta-Q-PN-expansion-Mink}
    \theta_Q 
    &=
    \partial_j \mathcal{Q}^{j}_{0} + \frac{1}{c^2} \left(\partial_t \mathcal{Q}^{0}_{1} + \partial_j \mathcal{Q}^{j}_{2}
    \right) \nonumber \\
    &+ \mathcal{O}(c^{-4})\,, \\
\label{eq:Qmu-umu-PN-expansion-Mink}
    \mathcal{Q}^{\alpha} a_{\alpha} 
    &= \mathcal{Q}^{j}_{0} D^N v_j + \mathcal{O}(c^{-2})\,.
\end{align}

The expansion of the conservation of the fluid current,
Eq.~\eqref{eq:j-mu-conservation} is
\begin{align}
    &
    D^N\rho 
    + 
    \theta^N\rho 
    \nonumber\\
    &
    +
    \frac{1}{2c^2}\left(
        D^N\rho + \theta^N\rho
        +
        D^N\left(v^2\right)\rho
    \right)
    + 
    \mathcal{O}\left(c^{-4}\right)
    \nonumber\\
    =&
    D^N\rho 
    + 
    \theta^N\rho 
    +
    \frac{1}{2c^2}D^N\left(v^2\right)\rho
    + 
    \mathcal{O}\left(c^{-4}\right)=0
    .
\end{align}

We can use the above results to obtain the non-relativistic expansion
of the energy equation \eqref{eq:energy_equation_general}.
Inserting Eqs.~\eqref{eq:PN_expansion_E_mink}, \eqref{eq:PN_expansion_P_mink}, 
\eqref{eq:Q-0-PN-Mink-2-def}, \eqref{eq:Q-j-PN-Mink-2-def}, 
\eqref{eq:theta-Q-PN-expansion-Mink}, \eqref{eq:Qmu-umu-PN-expansion-Mink},
and \eqref{eq:j-mu-conservation}
into Eq.~\eqref{eq:energy_equation_general}, we find that
\begin{align}
\label{eq:PN-Minkowski-expansion-energy-equation}
    & \partial_j \mathcal{Q}^j_{0} +
    \frac{1}{c^2} \left[ D^N \mathcal{E}_2 + \left(\mathcal{E}_2 + \mathcal{P}_0 \right) \theta^N
    \nonumber \right. \\
    &+
    \left.
    \partial_t \mathcal{Q}^0_1 + \partial_j \mathcal{Q}^j_2
    +
    \mathcal{Q}^j_0 D^N v_j 
    \right.
    \nonumber\\
    &-
    \left.
    2\eta^{(0)} \sigma^N_{ij}\left(\sigma^N\right)^{ij}
    \right] + \mathcal{O}(c^{-4}) =0\,,
\end{align}
where, $\mathcal{Q}_0^j$ and $\mathcal{Q}^j_2$ are defined in Eq.~\eqref{eq:Q-j-PN-Mink-2-def}, $\mathcal{Q}^0_1$ is defined in Eq.~\eqref{eq:Q-0-PN-Mink-2-def} and
\begin{align}
    \mathcal{E}_2 
    \equiv& 
    \rho\epsilon 
    +
    \tau_{\epsilon,1}^{(0)} D^N \epsilon
    +
    \tau_{\epsilon,2}^{(0)} \theta^N
    +
    \tau_{\epsilon,3}^{(0)} D^N \left(\frac{m_b}{T}\right)  \,,\\
    \mathcal{P}_0 
    \equiv&
    p 
    + \tau_{p,1}^{(0)} D^N \epsilon
    + (\tau_{p,2}^{(0)} - \zeta^{(0)}) \theta^N
    \nonumber\\
    &
    + \tau_{p,3}^{(0)} D^N \left( \frac{m_b}{T}\right) \,.
\end{align}
We can further rewrite $\mathcal{E}_2$ and $\mathcal{P}_0$, 
by noting that $T\left(\epsilon,\rho\right)$.
We then have 
\begin{align}
    D^N\left(\frac{m_b}{T}\right)
    =&
    \partial_{\epsilon}\left(\frac{m_b}{T}\right)D^N\epsilon
    +
    \partial_{\rho}\left(\frac{m_b}{T}\right)D^N\rho
    \nonumber\\
    =&
    \partial_{\epsilon}\left(\frac{m_b}{T}\right)D^N\epsilon
    -
    \partial_{\rho}\left(\frac{m_b}{T}\right)\theta^N\rho
    \nonumber\\
    &+
    \mathcal{O}\left(c^{-2}\right)
    . 
\end{align}
Similarly, the expansion of the spatial components of the momentum equation
Eq.~\eqref{eq:momentum_equation_general}, are
\begin{align}
    \label{eq:PN-Minkowski-expansion-momentum-equation}
    &
    \rho D^Nv^j
    +
    \partial^j\mathcal{P}_0
    \nonumber\\
    &
    -
    2\partial_k\left(\eta^{(0)}\left(\sigma^N\right)^{jk}\right)
    +
    \mathcal{O}\left(c^{-2}\right)
    =
    0
    .
\end{align}
In deriving this, we also used $\mathcal{Q}^j_0=0$, which follows from the
energy equation. For more discussion, see Sec.~\ref{subsec:NR-limit-general}.
\section{Characteristics and causality of the non-relativistic equations of motion}\label{appendix:causality-general-NR}
The symbol of the perturbed system [Eq.~\eqref{eq:perturbed-equation-general}] is obtained by replacing $\partial_t \to b$ and $\partial_j \to \xi_j$
\begin{align}
\label{eq:full-symbol-nr}
    M = 
    \begin{pmatrix}
        0 &b^2 & \rho b \xi_l \\
        \tau_{\epsilon,\parallel}^{(0)} b^2 + \rho b + \tau_{\epsilon,\perp} \xi^2 & \tau_{\rho,\perp}^{(0)} \xi^2 & \tau_u^{(0)} b \xi_l  + p \xi_l\\
        \tau_{\epsilon v}^{(-2)} b \xi_j  + \partial_{\epsilon} p \xi_j & \partial_{\rho}p \xi_j & D^j_l + \rho b \delta^j_l
    \end{pmatrix}
\end{align}
where $\xi^2 \equiv \xi^j \xi_j$, and  
\begin{align}
    D^j_l 
    \equiv 
    -
    \left(\zeta^{(0)} + \frac{\eta^{(0)}}{3}  \right) \xi^j \xi_l 
    - 
    \eta \xi^2 \delta^j_l
    \,.
\end{align}
We've added and additional factor of $b$ ($\partial_t$) so as to use the formula given in Eq.~\eqref{eq:Z-det}. This does not affect the characteristic equation~\cite{Bemfica:2020zjp}.
The principal part of the symbol is given by
\begin{align}\label{eq:symbol-full-newt-general}
    \mathcal{M} = 
    \begin{pmatrix}
        0 &b^2 & \rho b \xi_l \\
        \tau_{\epsilon,\parallel}^{(0)} b^2 + \tau_{\epsilon,\perp}^{(0)} \xi^2 & \tau_{\rho,\perp}^{(0)} \xi^2 & \tau_u^{(0)} b \xi_l  \\
        \tau_{\epsilon v}^{(-2)} b \xi_j  & 0 & D^j_l 
    \end{pmatrix}
\end{align}
The determinant is given by [Eq.~\eqref{eq:Z-det}]
\begin{align}
    &\text{det} \mathcal{M} 
    = 
    \nonumber \\
    &b^2 \left(\eta \xi^2 \right)^3 \xi^2
    \Bigg[
        -
        b^2 
        \left( 
            \tau_{\epsilon,\parallel}^{(0)} \left( \zeta^{(0)} + \frac{4\eta^{(0)}}{3}\right) 
            + 
            \tau_{\epsilon v}^{(-2)} \tau_{u}^{(0)}
        \right) 
        \nonumber \\
        &
        \qquad 
        + 
        \xi^2 \left(
            -
            \tau_{\epsilon,\perp}^{(0)} \left( \zeta^{(0)} + \frac{4\eta^{(0)}}{3}\right) 
            + 
            \tau_{\epsilon v}^{(-2)} \tau_{\rho,\perp}^{(0)} \rho 
        \right) 
    \Bigg]
    .
\end{align}
There are only two real solutions to this characteristic equation,
which indicates the system is not hyperbolic.
The characteristic speeds of the two hyperbolic modes are
\begin{align}\label{eq:speed-energy-mode-general}
    c_{0}^2 
    &= 
    \frac{b^2}{\xi^2} 
    \equiv
    \frac{
        -
        \tau_{\epsilon,\perp}^{(0)} \left( \zeta^{(0)} + \frac{4\eta^{(0)}}{3}\right) 
        + 
        \tau_{\epsilon v}^{(-2)} \tau_{\rho,\perp}^{(0)} \rho 
    }{
        \tau_{\epsilon,\parallel}^{(0)} \left( \zeta^{(0)} + \frac{4\eta^{(0)}}{3}\right) 
        + 
        \tau_{\epsilon v}^{(-2)} \tau_{u}^{(0)}
    }
    .
\end{align} 
This is real provided that
\begin{widetext}
\begin{align}\label{eq:condition-real-speed-NR}
    \left[
    -\tau_{\epsilon,\perp}^{(0)} \left( \zeta^{(0)} + \frac{4\eta^{(0)}}{3}\right) + \tau_{\epsilon v}^{(-2)} \tau_{\rho,\perp}^{(0)} \rho 
    \right]
    \left[ 
    \tau_{\epsilon,\parallel}^{(0)} \left( \zeta^{(0)} + \frac{4\eta^{(0)}}{3}\right) + \tau_{\epsilon v}^{(-2)} \tau_{u}^{(0)}
    \right] > 0
\end{align}
We now show that this inequality holds provided the relativistic theory is causal.
To see this we expand the following causality relations
\begin{align}
   \label{eq:inequality-a2-geq-0-general}
   &
   a_2 > 0 \implies 
   \tau_{\epsilon,\parallel}^{(0)} \left( \zeta^{(0)} + \frac{4\eta^{(0)}}{3}\right) 
   + 
   \tau_{\epsilon v}^{(-2)} \tau_{u}^{(0)} 
   + 
   \mathcal{O}(c^{-2}) > 0
   \,, \\
   \label{eq:inequality-a3-geq-0-general}
   &
   a_3 >0 \implies 
   -
   \tau_{\epsilon,\perp}^{(0)} \left( \zeta^{(0)} + \frac{4\eta^{(0)}}{3}\right) 
   + 
   \tau_{\epsilon v}^{(-2)} \tau_{\rho,\perp}^{(0)} \rho 
   + 
   \mathcal{O}(c^{-2}) > 0
   \,.
\end{align}
\end{widetext}
These two equations can be combined to show that Eq.~\eqref{eq:condition-real-speed-NR} 
is satisfied, which implies that $c_0^2$ is real.

In fact, one can carry out the analysis with the full symbol [Eq.~\eqref{eq:full-symbol-nr}] 
and show that there is a diffusive shear mode with frequency
\begin{align}
    b = i \omega = \frac{\eta^{(0)}}{\rho} \xi^2
    .
\end{align}
This shows that the first-order relativistic fluid models are of Type I.
We also note that there is a non-hydrodynamic mode 
(that is, a mode $\omega$ such that $\lim_{\xi\to0}\omega\neq0$) with frequency
\begin{align}
    b = i \omega = -\frac{\rho}{\tau_{\epsilon,\parallel}^{(0)}}\,,
\end{align}
indicating that a requirement for the system to be stable is $\tau_{\epsilon,\parallel}^{(0)} > 0$.
\section{Determinant Identity}
For the convenience of the reader we quote a useful identity for the determinant of 
matrices which take the following form
\begin{align}
    Z = 
    \begin{pmatrix}
    0 & b^2 &  a_{13} \xi_{\nu} \\
    a_{21} & a_{22} & a_{23} \xi_{\nu} \\
    a_{31} v^{\mu} & a_{32} v^{\mu} & v^{\mu} \xi_{\nu} A + B \delta^{\mu}_{\nu}\,.
    \end{pmatrix}
\end{align}
By performing row reductions, we have 
\begin{widetext}
\begin{align}
    \det Z
    &=    
    \det \begin{pmatrix}
    0 & b^2 &  a_{13} \xi_{\nu} \\
    a_{21} & 0 & \left(a_{23} - \frac{a_{22}a_{13}}{b^2}\right)\xi_{\nu} \\
    a_{31} v^{\mu} & 0 & v^{\mu} \xi_{\nu} \left(A - \frac{a_{32}a_{13}}{b^2}\right)+ B \delta^{\mu}_{\nu}\,.
    \end{pmatrix}
    \nonumber\\
    &=
    - b^2a_{21} 
    \left[
        a_{21}\det\left(v^{\mu} \xi_{\nu} 
            \left[
                A 
                - 
                \frac{a_{32}a_{13}}{b^2} 
                - 
                \left(a_{23}-\frac{a_{22}a_{13}}{b^2}\right)\frac{a_{31}}{a_{21}}
            \right]
            + 
            B \delta^{\mu}_{\nu}
        \right)
    \right].
\end{align}
Using Sylvester's identity 
\begin{align}
    \text{det}\left[\lambda_1 \delta^{\mu}_{\nu} + \lambda_2 \xi_{\nu} v^{\mu}\right]
    =&
    \lambda_1^n\text{det}\left[
        \delta^{\mu}_{\nu} 
        + 
        \frac{\lambda_2}{\lambda_1} \xi_{\nu} v^{\mu}
    \right]
    \nonumber\\
    =&
    \lambda_1^{n-1} \left[\lambda_1 + \lambda_2 \xi^{\mu} v_{\mu} \right]\,,
\end{align}
where, $n = \text{dim} [\xi_{\nu}]$, we see that
\begin{align}\label{eq:Z-det}
    &
    \text{det} Z 
    = 
    - 
    B^{n-1} \left\{  
        B b^2 a_{21} 
        + 
        v^2 
        \left[
            b^2 \left( A a_{21} - a_{31} a_{23} \right) 
            + 
            a_{13} \left(a_{22} a_{31} - a_{32} a_{21} \right) 
        \right] 
    \right\}
    .
\end{align} 
\end{widetext}

\bibliography{ref}
\end{document}